\documentstyle[aps,multicol,eqsecnum,epsf]{revtex}
\begin{document}
\draft
\title{ Scaling properties of driven interfaces in disordered media }

\author{Lu\'{\i}s A. Nunes Amaral, Albert-L\'aszl\'o
Barab\'asi\cite{byline}, Hern\'an A. Makse, and H. Eugene Stanley}

\address{ Center for Polymer Studies and Dept. of Physics,
Boston University, Boston, Massachussetts 02215 }

\date{\today}

\maketitle

\begin{abstract}

  We perform a systematic study of several models that have been
proposed for the purpose of understanding the motion of driven
interfaces in disordered media.  We identify two distinct universality
classes: (i) One of these, referred to as directed percolation
depinning (DPD), can be described by a Langevin equation similar to
the Kardar-Parisi-Zhang equation, but with quenched disorder.  (ii)
The other, referred to as quenched Edwards-Wilkinson (QEW), can be
described by a Langevin equation similar to the Edwards-Wilkinson
equation but with quenched disorder.  We find that for the DPD
universality class the coefficient $\lambda$ of the nonlinear term
diverges at the depinning transition, while for the QEW universality
class either $\lambda = 0$ or $\lambda \to 0$ as the depinning
transition is approached.  The identification of the two universality
classes allows us to better understand many of the results previously
obtained experimentally and numerically.  However, we find that some
results cannot be understood in terms of the exponents obtained for
the two universality classes {\it at\/} the depinning transition.  In
order to understand these remaining disagreements, we investigate the
scaling properties of models in each of the two universality classes
{\it above\/} the depinning transition.  For the DPD universality
class, we find for the roughness exponent $\alpha_P = 0.63 \pm 0.03$
for the pinned phase, and $\alpha_M = 0.75 \pm 0.05$ for the moving
phase.  For the growth exponent, we find $\beta_P = 0.67 \pm 0.05$ for
the pinned phase, and $\beta_M = 0.74 \pm 0.06$ for the moving phase.
Furthermore, we find an anomalous scaling of the prefactor of the
width on the driving force.  A new exponent $\varphi_M = -0.12 \pm
0.06$, characterizing the scaling of this prefactor, is shown to
relate the values of the roughness exponents on both sides of the
depinning transition.  For the QEW universality class, we find that
$\alpha_P \approx \alpha_M = 0.92 \pm 0.04$ and $\beta_P \approx
\beta_M = 0.86 \pm 0.03$ are roughly the same for both the pinned and
moving phases.  Moreover, we again find a dependence of the prefactor
of the width on the driving force.  For this universality class, the
exponent $\varphi_M = 0.44 \pm 0.05$ is found to relate the different
values of the local $\alpha_P$ and global roughness exponent $\alpha_G
\approx 1.23 \pm 0.04$ at the depinning transition.  These results
provide us with a more consistent understanding of the scaling
properties of the two universality classes, both at and above the
depinning transition.  We compare our results with all the relevant
experiments.

\end{abstract}


\begin{multicols}{2}

\section{Introduction}

  Recently the growth of rough interfaces has witnessed an explosion
of theoretical, numerical, and experimental studies, fueled by the
broad interdisciplinary aspects of the subject
\cite{vfk,krug,meakin,hhz,alb,grs}.  Applications can be so diverse as
imbibition in porous media, fluid--fluid displacement, fire front
motion, and the motion of flux lines in superconductors
\cite{stokes,rubio,horv,wong,zzhang,Havlin,Buldyrev,Buldyrev1,Buldyrev2,amaral1,amaral2}.

  In general, a $d$-dimensional self-affine interface, described by a
single-valued function $h({\bf x},t)$, evolves in a
$(d+1)$-dimensional medium.  Usually, some form of disorder $\eta$
affects the motion of the interface leading to its roughening.  Two
main classes of disorder have been discussed in the literature.  The
first, called thermal or ``annealed,'' depends only on time.  The
second, referred to as ``quenched,'' is frozen in the medium.  Early
studies focused on {\it time-dependent\/} disorder as being
responsible for the roughening
\cite{vis,family,mrsc,kk,ew,kpz,medina,Zhang,Barabasi}.  Here, we
consider in detail the effect of {\it quenched\/} disorder on the
growth.

  The presence of quenched disorder introduces an interesting analogy
between the motion of driven interfaces in disordered media and the
theory of critical phenomena.  The continual motion of the interface
requires the application of a driving force $F$.  There exists a
critical force $F_c$, such that for $F < F_c$, the interface will
become pinned by the disorder after some finite time.  For $F > F_c$,
the interface moves indefinitely with an average velocity $v(F)$.
This suggests that the motion of driven rough interfaces in disordered
media can be studied as a phase transition --- which we shall call the
{\it depinning transition}.  The velocity of the interface $v$ plays
the role of the {\it order parameter}, since as $F \rightarrow F_c^+$,
$v$ vanishes as
\begin{equation}
v \sim f^{\theta}.
\label{vel}
\end{equation}
We call $\theta$ the {\it velocity\/} exponent, and $f \equiv (F-F_c)
/ F_c$ the {\it reduced force} (Fig. \ref{phases}).

  For $F \rightarrow F_c^+$, large but finite regions of the
interface are pinned by the disorder.  At the transition, the
characteristic length $\xi$ of these pinned regions diverges as
\begin{equation}
\xi \sim f^{-\nu},
\label{corl}
\end{equation}
where $\nu$ is the {\it correlation length\/} exponent.

  The added phenomenological richness introduced by the quenched
disorder leads to an increased difficulty.  In fact, the problem of
identifying the universality classes for interface roughening in the
presence of quenched disorder and determining the corresponding sets
of critical exponents has so far remain unsolved.  Experimental and
numerical measurements of some of the critical exponents vary
considerably, and the question of the existence of universality
classes bas been raised
\cite{stokes,rubio,horv,wong,zzhang,Havlin,Buldyrev,Buldyrev1,Buldyrev2,amaral1,amaral2,robbins1,robbins2,kessler,koplik,parisi,amaral3,Tang,lesc,dong,lesc1,jullien,alstrom,Roux94a,Galluccio95a,csahok,amaral4,Tang95,havlin2,cayley,makse2,makse}.
The problem is made more complex by the disagreement of most
experimentally measured values with analytical calculations
\cite{grinstein,theory1,theory2}.

  In this paper, we address the question of why the same exponents
vary in value for different systems.  We study several numerical
models introduced to study interface roughening in the presence of
quenched disorder, and identify two universality classes.  We show
that one of those universality classes can be identified with the case
studied analytically.  The exponents for the other universality class
cannot be determined from renormalization groups calculations.
However, mappings to directed percolation (DP) and isotropic
percolation allow us to estimate most scaling exponents.

  The identification of the universality classes and the calculation
of the respective scaling exponents {\it at\/} the depinning
transition enable us to understand the results for most of the models
proposed and for some of the experiments.  We shall see that several
experimental results --- particularly for the case of the moving
interface in fluid-fluid displacement experiments --- still do not fit
the framework provided by the universality classes we discuss.  For
this reason, we study models in each of the universality classes {\it
above\/} the depinning transition --- i.e., in the {\it moving\/}
phase.  The results obtained allow us to re-interpret the experimental
measurements.  Furthermore, we find that for one of the universality
classes, some of the critical exponents change values at the depinning
transition.

  The paper is organized as follows:  In Sec. II we introduce the
problem, by describing the set of relevant scaling properties and
associated exponents, for driven interfaces moving in disordered
media.  In Sec.  III we discuss the experimental and numerical results
motivated our study.  In Sec. IV we consider several numerical models,
and identify two distinct universality classes.  In Sec. V we try to
link the results obtained for the two universality classes with the
experimental results.  To achieve this goal, we study models,
representative of each of the universality classes, above the
depinning transition.  Finally, in Sec.  VI, we summarize our results
and discuss some of the new questions arising from this work.

\section{Scaling Properties and Critical Exponents}

  An interface moving in a disordered medium becomes rough due to the
action of the disorder.  This roughening process can be quantified by
studying the {\it global\/} interface width
\begin{equation}
W(L,t) \equiv \left \langle \left [ \overline{h^2({\bf x},t)} -
\overline{h({\bf x},t)}^2 \right ] ^{1/2} \right \rangle,
\label{wid}
\end{equation}
where $L$ is the system size, the overbar denotes a spatial average, and
the angular brackets denote an average over realizations of the disorder.

  The study of discrete models \cite{vis,family,mrsc,kk} and continuum
growth equations \cite{ew,kpz} leads to the observation that during
the initial period of the growth, i.e. for $t \ll t_{\times}(L)$, the
global width $W$ grows with time as
\begin{equation}
W(t) \sim t^{\beta} \qquad (t \ll t_{\times}),
\label{bet}
\end{equation}
where $\beta$ is the {\it growth\/} exponent.  For times much larger
than $t_{\times}$ the width saturates to a constant value.  It was
observed that the saturation width of the interface $W_{\rm sat}$
scales with $L$ as
\begin{equation}
W_{\rm sat} \sim L^{\alpha} \qquad (t \gg t_{\times}),
\label{wsat}
\end{equation}
where $\alpha$ is the {\it roughness\/} exponent.

  The dependence of $t_{\times}$ on $L$ allows the combination of
(\ref{bet}) and (\ref{wsat}) into a single scaling law \cite{vis}
\begin{mathletters}
\begin{equation}
W(L,t) \sim L^{\alpha}~ f( t / t_{\times} )
\end{equation}
where
\begin{equation}
t_{\times} \sim L^{z}.
\end{equation}
\label{scl1}
\end{mathletters}
\noindent
Here $z = \alpha / \beta$ is the {\it dynamical\/} exponent, and
$f(u)$ is a universal scaling function that grows as $u^{\beta}$
when $u \ll 1$, and approaches a constant when $u \gg 1$.

  An alternative way of determining the scaling exponents is to study
the {\it local\/} width $w$ in a box of length $\ell < L$.  The
scaling law (\ref{scl1}), and the fact that the interface
is self-affine (and with $\alpha < 1$), allow us to conclude
\begin{mathletters}
\begin{equation}
w(\ell,t) \sim \ell^{\alpha}~ f( \ell / \ell_{\times} ),
\end{equation}
where
\begin{equation}
\ell_{\times} \sim t^{1/z} \qquad (t \ll t_{\times}),
\end{equation}
or
\begin{equation}
\ell_{\times} \sim L \qquad (t \gg t_{\times}).
\end{equation}
\label{scl2}
\end{mathletters}
\noindent
Here $f(u)$ is a universal scaling function that decreases as
$u^{-\alpha}$ when $u \gg 1$ and approaches a constant when $u \ll 1$.

  Not all of the exponents introduced so far are independent; e.g., we
can calculate the velocity exponent from the other scaling exponents,
as we show by the following argument. Near the depinning transition,
except for a few growing regions, most of the interface is pinned by
some pinning path.  The growth occurs by the propagation of these
growing regions through the system.  The characteristic time required
for this propagation must be of the order of $t_{\times}$, since this
is the time for correlations to propagate across the entire system, as
defined by (\ref{scl1}b).  During this process the interface advances
from one blocking path to the next, the distance advanced is typically
of the order of $W_{\rm sat}$.  Since, close to the transition, $\xi
\sim L$, we can use (\ref{corl}), (\ref{wsat}), and (\ref{scl1}) to
obtain
\begin{equation}
v \sim W_{\rm sat}~/~t_{\times} \sim \xi^{\alpha}~/~\xi^{z}
\sim \xi^{\alpha - z} \sim
f^{-\nu (\alpha - z)}.
\label{eq101}
\end{equation}
Upon comparison with (\ref{vel}), we conclude that
\begin{equation}
\theta = \nu (z - \alpha).
\label{thet}
\end{equation}

  This exponent ``scaling law'' can also be derived in a different way
\cite{theory1,theory2}, by considering the interplay between the range
of action of the quenched disorder and the length scales for which the
disorder can be considered to be time-dependent because of the motion
of the interface.  The quenched disorder is a function of ${\bf x}$
and $h$, but we can consider the change of variables to a reference
frame moving with the velocity of the interface
\begin{equation}
\eta(x,h) \to \eta(x, vt+h).
\end{equation}
Now, we can consider the conditions for which $vt >> h$.  This will
happen for the length scale characterizing the action of the quenched
disorder, $\xi$.  Thus we have
\begin{equation}
v \sim h / t \sim \xi^{\alpha} / \xi^z \sim \xi^{\alpha - z}
\sim f^{\nu (z - \alpha)},
\end{equation}
leading to (\ref{thet}).

\section{Driven Interfaces in Disordered Media}

\subsection{Experiments}

  The most commonly measured exponent in experiments and simulations
is the roughness exponent $\alpha$.  In the last few years, several
experiments in which quenched disorder plays a dominant role have been
performed.

{\it (i)\/} Stokes {\it et al.} performed fluid-fluid displacement
experiments, and observed that for some conditions a self-affine
interface would form, while for other, the growth would generate a
percolation-like cluster \cite{stokes}.

{\it (ii)\/} Rubio {\it et al.} measured $\alpha$ for fluid-fluid
displacement experiments \cite{rubio}, consistently obtaining $0.73$
regardless of the velocity of the interface $v$.  Moreover, they
observed that the total width of the interface decreased with $v$
\cite{rubio}.

{\it (iii)\/} Horv\'ath {\it et al.} performed similar fluid-fluid
displacement experiment, obtaining $\alpha \approx 0.81$ \cite{horv}.

{\it (iv)\/} Buldyrev {\it et al.} studied the imbibition of coffee
in paper towels observing $\alpha \approx 0.63$ for the pinned
interface \cite{Havlin,Buldyrev,amaral1,amaral2}.  Similar experiments
were performed by Buldyrev {\it et al.} for $2+1$ dimensions
leading to $\alpha \approx 0.52$ \cite{Buldyrev1}.

{\it (v)\/} He {\it et al.} conducted fluid-fluid displacements
experiments for different velocities of the interface \cite{wong}.
They reported values of $\alpha$ varying from $0.6$ for high
velocities, to $0.9$ for low velocities.  Furthermore, they also
observed that the total width of the interface decrease with velocity
\cite{wong}.

{\it (vi)\/} Zhang {\it et al.} performed flameless paper burning
experiments \cite{zzhang}.  They reported $\alpha \approx 0.71$ for the
moving interface.

  Thus the experiments reveal values of the roughness exponent as low
as $0.6$ and as high as $0.9$, leading many to question whether one
can divide all systems into a small number of universality classes
(Fig. \ref{f:summary}).

\subsection{Discrete models}

  Several discrete models with quenched disorder have been proposed to
explain the experimental results.

{\it (i)\/} Early on, Cieplak and Robbins noticed the importance of
quenched disorder in many surface phenomena \cite{robbins1}.  They
studied two models, motivated by fluid-fluid displacement experiments.
The first, called the {\it fluid invasion model}, is a quite realistic
model and calculations reveal $\alpha \approx 0.8$ in $1+1$ dimensions
\cite{robbins1,robbins2}.  The second, called the {\it random field
Ising model\/} (RFIM) \cite{grinstein}, does not have a self-affine
phase in $1+1$ dimensions, so no roughness exponent can be calculated.
On the other hand, for $2+1$ dimensions, a self-affine regime exists,
and a roughness exponent of $0.66$ was calculated
\cite{robbins1,robbins2}.

{\it (ii)\/} Kessler {\it et al.} integrated the Edwards-Wilkinson
(EW) equation with quenched disorder, with the purpose of explaining
the results for the fluid-fluid displacement experiments
\cite{kessler}.  They reported values of $\alpha$ that vary with the
driving force, i.e. with the velocity of the interface.  For low
velocities they found $\alpha \approx 1$, and for large velocities,
$\alpha \approx 0.5$ \cite{kessler}.  Furthermore, they found a
decrease of the total width with the velocity of the interface.
Similar results were obtained earlier by Koplik and Levine
\cite{koplik}.

{\it (iii)\/} Buldyrev {\it et al.} introduced a discrete model to
explain their imbibition experiments \cite{Havlin,Buldyrev,amaral2}.
They found that in $1+1$ dimensions the scaling behavior of the pinned
interface can be obtained exactly by mapping the interface onto DP.
In higher dimensions they found that the interface can be mapped onto
a {\it directed surface\/} (DS) \cite{Buldyrev1}.  In $1+1$
dimensions, DP and DS are equivalent, so this model is referred to as
{\it directed percolation depinning\/} (DPD) model
\cite{Buldyrev1,amaral2}.

{\it (iv)\/} A similar model was introduced independently by Tang and
Leschhorn \cite{Tang}.  For the DPD class of models, it was observed
that $\alpha \approx 0.63$ for the pinned interface
\cite{Havlin,Buldyrev,Tang}.

{\it (v)\/} Parisi also introduced a model for interface roughening;
however, no measurement was made of the roughness exponent
\cite{parisi}.  Several authors studied Parisi's model and found it to
possess problems (see Appendix B).  For this reason, Amaral introduced
an alternate version of the model, which led to exponents identical
to the ones obtained for the DPD class of models \cite{amaral3}.

{\it (vi)\/} Dong {\it et al.} also integrated numerically the QEW
equation, and reported $\alpha \approx 0.97$ from the study of the
height-height correlation function \cite{dong}.

{\it (vii)\/} Leschhorn introduced a model, corresponding essentially
to a discretization of the QEW equation, and found $\alpha \approx
1.25$ from measurements of the global width.  The disagreement with
Ref. \cite{dong} was later explained by Leschhorn and Tang, by showing
that the height-height correlation function is limited to a scaling
with a roughness exponent smaller or equal to $1$ \cite{lesc1}.

{\it (viii)\/} Roux and Hansen studied what amounts to a
self-organized version of Leschhorn's model finding that $\alpha
\approx 0.86$ for the local width and $\alpha \approx 1.22$ for the
global width \cite{Roux94a}.  Using a similar model, Galluccio and
Zhang found identical results \cite{Galluccio95a}.

{\it (x)\/} Csah\'ok {\it et al.} studied the effect of multiplicative
quenched disorder in the scaling properties of the interface
\cite{csahok}.  They found $\beta \approx 0.65$ and $\alpha \approx
0.8$, in close agreement with the experimental results of Horv\'ath
{\it et al.} \cite{horv}.

\subsection{Continuum models}

  On the continuum front, some attempts were made to explain the
previous results.  The first theoretical studies to introduce quenched
disorder were performed independently by Nattermann {\it et al.}
\cite{theory1} and Narayan and Fisher \cite{theory2}.  Analyzing the
Kardar-Parisi-Zhang (KPZ) equation \cite{kpz} with quenched disorder,
they argued that the dependence of the noise on $h({\bf x}, t)$
introduces an infinite hierarchy of nonlinearities.  They also noticed
that in the annealed disorder case the KPZ nonlinearity has a kinematic
origin.  Thus $\lambda \sim v$, and $\lambda$ should vanish at the
depinning transition.  For these reasons, they argue that at the
depinning transition, it is sufficient to consider the equation
\cite{grinstein,bruins}
\begin{equation}
\frac{\partial h}{\partial t} = F + \eta({\bf x},h) + \nabla^2 h,
\label{qew}
\end{equation}
where $\eta({\bf x},h)$ represents the quenched disorder.  This
equation was studied by means of the functional renormalization group
\cite{grinstein,theory1,theory2}, yielding to first order
\begin{mathletters}
\begin{equation}
\alpha = \epsilon / 3,
\end{equation}
\begin{equation}
\nu = 1 / (2 - \alpha),
\end{equation}
and
\begin{equation}
z = 2 - 2 \epsilon / 9,
\end{equation}
\label{theory}
\end{mathletters}
where $\epsilon \equiv 4 - d$.  Hence $\alpha = 1$ in $1+1$
dimensions, in disagreement with many of the experimental and numerical
results. Reference \cite{theory2} also argued that all but
(\ref{theory}c) are exact for all orders in $\epsilon$.

  In view of the disagreement between the theoretical predictions of the
EW equation with quenched disorder and the results presented in the
previous subsections, Csah\'ok {\it et al.} numerically integrated the
KPZ equation with quenched disorder
\cite{csahok}
\begin{equation}
\frac{\partial h}{\partial t} = F + \eta({\bf x},h) + \nabla^2 h
+ \lambda (\nabla h)^2.
\label{qkpz}
\end{equation}
The numerical integration of (\ref{qkpz}) for the {\it moving\/} phase
\cite{csahok} yielded exponents in agreement with the calculations for
the models in the DPD universality class.

\section{The Depinning Transition}

\subsection{Looking for nonlinearities}

  The fact that for (\ref{qew}) we can calculate the exponents, and they
do not agree with the numerical results obtained for (\ref{qkpz}),
suggests that the nonlinear term $\lambda (\nabla h)^2$ may play a
crucial role in determining the scaling properties of the interface.
Unfortunately no analytical results are available for Eq. (\ref{qkpz}).
In this section, we investigate the presence of the nonlinear term in
several of the models discussed in the previous section \cite{amaral4}.

  If we consider an equation of the KPZ-type, and consider its average
over the spatial coordinates, we obtain
\begin{equation}
v = \overline{\partial h / \partial t} = \overline{F + \eta +
\nabla^2 h + \lambda (\nabla h)^2}.
\label{ave_v}
\end{equation}
Now, if we impose an average {\it tilt\/} to the interface, $m \equiv
\overline{\nabla y}$, and make the transformation
\begin{equation}
h(x,t) \rightarrow y(x,t) \equiv h(x,t) + mx ,
\end{equation}
we obtain
\begin{equation}
v(m) = \overline{\partial y / \partial t} = \overline{F + \eta +
\nabla^2 h + \lambda (\nabla h)^2} + \lambda m^2.
\end{equation}
So, it follows that the average velocity of the interface changes
with the tilt as \cite{Krug90a}
\begin{equation}
v(m) = v + \lambda m^2.
\label{vel1}
\end{equation}
Thus by varying the tilt $m$, we can study the presence of nonlinear
terms in the growth equation and calculate the effective coefficient
$\lambda$ of a given model \cite{Krug90a}.

  For discrete models the tilt can be easily implemented introducing
{\it helicoidal\/} boundary conditions.  In $1+1$ dimensions, we
simply impose the condition, $h(0,t) = h(L,t) - Lm$ and $h(L+1,t) =
h(1,t) + Lm$.  In $d+1$ dimensions, we use the same boundary
conditions, so that the tilt occurs only in one direction.

\subsection{The DPD universality class}

\subsubsection{The DPD-1 model}

  We begin by applying the method described above to the model
introduced by Tang and Leschhorn \cite{Tang}.  In $1+1$ dimensions, the
interface becomes pinned by a DP cluster \cite{Buldyrev,Tang}, and the
critical dynamics is controlled by a divergent correlation length
parallel to the interface $\xi \sim f^{-\nu}$, with $\nu \approx 1.73$.
This model, referred to as ``DPD-1'', excludes overhangs, and gives
rise to a self-affine interface at the depinning transition, with a
roughness exponent $\alpha \approx 0.63$.

  In the DPD-1 model, in $1+1$ dimensions, we start from a horizontal
interface at the bottom edge of a lattice of size $L$.  At every site
of the lattice we define a random, uncorrelated quenched variable, the
noise $\eta$, with magnitude in the range $[0,1]$. During the time
evolution of the interface, we choose one of the $L$ columns at
random \cite{expl0}.  If the difference in height to the lowest neighbor
is larger than $+1$, this lowest neighboring column grows by one unit.
Otherwise, the chosen column grows one unit provided the noise on the
site above the interface is smaller than the driving force $F$.  The
unit time is defined to be $L$ growth attempts.

  We measure the velocity of the interface for different values of the
reduced force $f$ and different values of the tilt $m$.  The results
for $1+1$ dimensions are shown in Fig. \ref{dpd1}(a).  For a fixed
force $f$, we find that the interface velocity depends on the tilt
$m$, indicating the existence of nonlinear terms.  Near the depinning
transition ($f \rightarrow 0$), the velocity curves become ``steeper''
and, from (\ref{vel1}), we infer that $\lambda$ is coupled to the
external force and that it increases as $f \rightarrow 0$.

  To measure $\lambda$, we first attempt to fit a parabola to the
tilt-dependent velocities in the vicinity of zero tilt. The
calculations indicate that as we approach the depinning transition,
$\lambda$ diverges as \cite{amaral4}
\begin{equation}
\lambda \sim f^{-\phi}.
\label{phi}
\end{equation}
However, in the vicinity of $F_c$, the velocity curves lose their
parabolic shape for large tilts (see Figs. \ref{dpd1}(a) and
\ref{cross}), indicating a change to a different universality class
\cite{Tang95}.

  We can understand the breakdown of (\ref{vel1}) for large $m$ using
scaling arguments.  Substituting Eqs. (\ref{vel}) and (\ref{phi}) into
(\ref{vel1}), we find \cite{amaral4}
\begin{equation}
v(m, f) \propto  f^{\theta} + a  f^{-\phi} m^2.
\label{velo}
\end{equation}
Equation (\ref{velo}) indicates that the velocity curves corresponding
to two different forces $f_1$ and $f_2$, with $f_1 > f_2$, will
intersect at a tilt $m_{\times}$ (see Fig. \ref{cross}). For tilts
greater than $m_{\times}$, $v(m, f_1) < v(m, f_2)$, a clearly unphysical
prediction since $f_1 > f_2$, and since the average velocity, for the
same tilt $m$, should be larger for the larger force.  Thus the velocity
cannot follow a parabola for arbitrarily large $m$, and a crossover to a
different behavior than that of Eq. (\ref{velo}) must occur for values
of the tilt larger than $m_{\times}$.

  Letting $(f_1 - f_2) \rightarrow 0$, we find from (\ref{velo}) that
the crossing point of the two corresponding parabolas scales as
\begin{equation}
m_{\times}^2 \sim f^{\theta + \phi}.
\label{range}
\end{equation}
Equations (\ref{velo}) and (\ref{range}) motivate the scaling form for
the velocities \cite{amaral4}
\begin{equation}
v(m, f) \sim f^{\theta} g(m^2 / f^{\theta + \phi}),
\label{scal}
\end{equation}
where $g(x) \sim ~$const for $x \ll 1$, and $g(x) \sim
x^{\theta/(\theta + \phi)}$ for $x \gg 1$ \cite{explain00}.  Figure
\ref{dpd1} shows the data collapse we obtain using (\ref{scal}),
with the exponents
\begin{equation}
\theta = 0.64 \pm 0.08, \quad  \phi= 0.64 \pm 0.08,
\end{equation}
for $1+1$ dimensions.  Below, we will show that an entire class of
models produces similar results, which will allow us to group them
into a single universality class.

\subsubsection{Other models}

  The scaling behavior (\ref{scal}) is not limited to the DPD-1 model
in $1+1$ dimensions.  For $2+1$ dimensions and for the models
introduced in Refs. \cite{Havlin,parisi,amaral3}, we find a very
similar behavior (Table \ref{tab1}).

  The ``DPD-2'' model was introduced by Buldyrev {\it et al.}
\cite{Havlin,Buldyrev,Buldyrev1} (see also Amaral {\it et al.}
\cite{amaral2}).  Let us define the model for $1+1$ dimensions.  In a
square lattice of edge $L$, assign an uncorrelated random number, the
disorder $\eta_i$, with magnitude uniformly distributed in the
interval $[0,1]$, to each cell $i$.  We compare the random pinning
forces $\eta_i$ in the lattice with the driving force $F$, where $0
\le F \le 1$.  If the pinning force at a certain cell $\eta_i$ is
larger than the driving force, the cell is labeled ``blocked'',
otherwise it is labeled ``unblocked''.  Thus a cell is blocked with a
probability $p = 1 - F$.

  Since the model was developed to study imbibition, we will refer to
the growing, invading, region as ``wet'', and to the invaded region as
``dry''.  At time $t=0$, we wet all cells in the bottom row of the
lattice.  Then, we select a column at random \cite{expl0}, and wet all
dry {\it unblocked cells\/} in that column that are nearest-neighbors
to a wet cell.  To obtain a single-valued interface, we impose the
auxiliary rule that all dry {\it blocked cells\/} below a wet cell
become wet as well \cite{explain0}.  We refer to this rule as {\it
erosion of overhangs}.  The time unit is defined as $L$ growth
attempts.

  Figure \ref{dpd2} shows the dependence of the velocity on the tilt
for the ``DPD-2'' model and the respective data collapse according to
(\ref{scal}), for $1+1$ dimensions.  The values of the exponents in
the data collapse are \cite{explain1}
\begin{equation}
\theta = 0.59 \pm 0.12, \quad \phi= 0.55 \pm 0.12.
\end{equation}

  Finally, we simulate the model proposed by Amaral \cite{amaral3},
which we will refer to as ``DPD-3''.  We simulate it in $1+1$
dimensions, and rescale the tilt dependent velocities according to
(\ref{scal}) using the exponents (Fig. \ref{dpd3})
\begin{equation}
\theta = 0.70 \pm 0.12, \quad \phi = 0.65 \pm 0.12.
\end{equation}

\subsubsection{Discussion}

  Three questions are raised by these results:

{\it (i)\/} Is the new exponent $\phi$ an independent exponent or can
it be expressed in terms of the other known exponents?

{\it (ii)\/} What are the scaling properties of the tilted interface?

{\it (iii)\/} What are the reasons for the divergence of $\lambda$ at
the depinning transition?

  Questions {\it (i)\/} and {\it (ii)\/} have recently been answered
by Tang, Kardar and Dhar (TKD) \cite{Tang95}.  They noticed that for
the DPD class of models there is a characteristic slope
\begin{equation}
m_c = \xi_{\perp} / \xi_{\parallel} \sim f^{ \nu (1 - \alpha)},
\label{slop}
\end{equation}
where $\xi_{\perp}$ and $\xi_{\parallel}$ are, respectively, the
perpendicular and parallel correlation lengths of a DP cluster (see
Appendix A).

  For values of the average tilt smaller than $m_c$, the interface
still belongs to the DPD universality class.  However, for larger
values of the tilt a qualitative change should occur.  But we showed
above that such a change occurs for $m > m_{\times}$, so we conclude
that $m_c = m_{\times}$, from which follows \cite{Tang95}
\begin{equation}
\theta + \phi = 2 \nu (1 - \alpha).
\label{e:t+p}
\end{equation}
Combining (\ref{e:t+p}) with (\ref{vel}), we find the scaling law
\begin{equation}
\phi = \nu (2 - \alpha - z).
\label{phi3}
\end{equation}

  TKD also showed that for $m > m_{\times}$ a {\it new\/} universality
class is obtained \cite{Tang95}, and they were able to relate its scaling
exponents to the exponents of the KPZ equation with annealed disorder
\cite{Tang95}
\begin{equation}
\alpha_d = {\alpha_{d-1}^{\rm KPZ}}/{z_{d-1}^{\rm KPZ}}, \qquad
z_d = {1}/{z_{d-1}^{\rm KPZ}}.
\end{equation}

  To address question {\it (iii)}, let us consider the effect of the
nonlinear term.  Its role is to make the interface grow wherever it has
a nonzero slope, so it corresponds to what we can refer to as lateral
growth.  When an interface is moving, we can consider two types of
growth, forward growth and lateral growth.  The meaning of $\lambda
\rightarrow \infty$ is then the irrelevance of forward growth compared
with lateral growth when we approach the depinning transition.  These
ideas were studied quantitatively by Makse \cite{makse3}, who found
that, near the depinning transition, the probability of occurrence of
lateral growth vanishes much slower than the probability of occurrence
of forward growth.  Furthermore, Makse argued that the application of
constraints to the local slopes of the interfaces, as is the case for
the DPD class of models, appears to be an {\it essential\/} ingredient
for $\lambda$ to diverge \cite{makse3}.

\subsection{The QEW universality class}

\subsubsection{Discretizations of the QEW equation}

  We mentioned in Sec. III that several authors numerically integrated
Eq. (\ref{qew}), and others developed models corresponding to the
discretization of this equation.  Thus, we expect results in agreement
with the theoretical calculations.  Furthermore, we do not expect, for
these models, to detect any dependence of the velocity on the tilt of
the interface.  In fact, we have simulated the model introduced by
Leschhorn \cite{lesc}, which we will refer to as ``QEW-1'', and find
that for any reduced force, $\lambda = 0$ (Fig. \ref{qew1}).

  A Hamiltonian model, which we will refer to as ``QEW-2'', was
introduced by Makse \cite{makse2}.  In $1+1$ dimensions, the
Hamiltonian is defined as
\begin{equation}
{\cal H} ~ = \sum_{i=1}^L \left [ (h_{i+1} - h_i )^2 - F h_i + \zeta(i,
h_i) \right ].
\label{hamilt}
\end{equation}
Here, the first term represents the elastic energy that tends to
smooth the interface, and $\zeta(i, h_i)$ is an uncorrelated random
number which mimics a random potential due to the disorder of the
medium.  In the simulation, a column $i$ is chosen and its height is
updated to $(h_i + 1)$ if the change in (\ref{hamilt}) is
negative. Thus, only motions that decrease the total energy of the
system are accepted.  Backward motions are neglected since these are
rare events.  Equation (\ref{qew}) can be easily derived from this
Hamiltonian simply by considering that
\begin{equation}
\frac{\partial h}{\partial t} = - \frac{\delta {\cal H}}{\delta h},
\qquad \eta = - \frac{\delta \zeta}{\delta h}.
\end{equation}

  We study the dependence of the velocity with the tilt for this model
and find that $\lambda = 0$ for any force, just as for the QEW-1
model.

\subsubsection{The RFIM}

  An interesting result, presented in Sec. III, is the reasonable
agreement between the exponents measured for the FIM and the RFIM
\cite{robbins1,robbins2} with the predictions of (\ref{qew}).  Since
these models are not simple discretizations of Eq. (\ref{qew}), it
is interesting to investigate the presence of any nonlinear term.

  For this reason, we study the RFIM in both $1+1$ and $2+1$
dimensions.  This model allows for overhangs, and for certain values
of its parameters can be mapped to the isotropic percolation problem
\cite{robbins2}.  In the RFIM, spins on a square lattice interact
through the Hamiltonian
\begin{equation}
{\cal H} \equiv - \sum_{\langle i,j\rangle} S_iS_j - \sum_i
[F + \zeta(i,h)]S_i,
\label{rfim}
\end{equation}
where $S_i=\pm1$, $F$ now denotes the external magnetic field, and
$\zeta$ is the time-independent local random field (i.e., quenched
noise) whose values are uniformly distributed in the interval
$[-\Delta,\Delta]$.  The strength of the quenched disorder is
characterized by the parameter $\Delta$.  At time zero, all spins are
``down''--- except those in the first row, which are initially ``up''.
The interface consists of all down spins that are nearest-neighbors to
an up spin.  During the time evolution of the system, we flip any down
spin that belongs to the interface and is ``unstable,'' i.e., whenever
the flip will lower the total energy of the system.  The control
parameter of the depinning transition is the external magnetic field
$F$; the unit time corresponds to the parallel flipping of all
unstable spins \cite{explain2}.

  For $1+1$ dimensions, there are two morphologically-different
regimes, depending on the strength $\Delta$ of the disorder (i.e., of
the random fields).  For $\Delta > 1.0$, the interface is self-similar
(SS), while for $\Delta < 1.0$ it is faceted or flat (FA).  For $2+1$
dimensions, there is again a FA regime ($\Delta < 2.4$), a SS regime
($\Delta > 3.4$), and also a self-affine (SA) regime in between ($2.4
< \Delta < 3.4$) \cite{robbins1}.  The SA regime, which exists only
for the case of $2+1$ dimensions, is the only regime of the RFIM for
which either Eqs.  (\ref{qew}) or (\ref{qkpz}) could apply.  In the SS
regime, overhangs in the interface are larger enough for the
approximation of small slopes not to be valid.  On the other hand, in
the FA regime, lattice effects dominate the growth \cite{expl-fa}.

  For the SA and SS regimes we find that (\ref{scal}) is still valid;
however, we find a negative $\phi$
\begin{equation}
\lambda \sim f^{| \phi |} \rightarrow 0.
\label{phi2}
\end{equation}
This behavior can be understood, for the SS regime, by considering
that near the depinning transition, the morphology of the interface
corresponds to the hull of an isotropic percolation cluster, which has
no well-defined orientation \cite{robbins1}.  Thus a change in the
boundary conditions will not affect the growth process, and we cannot
expect any divergence of a possible nonlinear term when the magnetic
field approaches its critical value.  On the other hand, for large
fields the effect of the quenched disorder diminishes, and we can
observe an average interface orientation.  For such values of field,
we expect the presence of nonlinear terms, of {\it kinematic\/}
origin, to be relevant.  Although for the SA regime the behavior of
$\lambda$ is similar (Fig. \ref{rfim2}), we do not have a simple
argument that would explain the observed behavior.

  These results lead us to conclude that in the SA regime the RFIM
belongs to the universality class of Eq.  (\ref{qew}).  This conclusion
is further supported by the agreement between the numerically
determined exponents, $\alpha \approx 0.67$ and $\theta \approx 0.60$
for $2+1$ dimensions, and the theoretical predictions of (\ref{qew}).

\subsection{Discussion}

  The results of Table \ref{tab1} show, for $1+1$ dimensions, a
separation into two groups in the values of the critical exponents for
the six models studied \cite{explain1}.  This separation reflects the
existence of two distinct universality classes, described by the two
continuum growth equations, (\ref{qew}) and (\ref{qkpz}).  For the
QEW-1, QEW-2 models, and for the RFIM in the SA regime, we find that
either $\lambda = 0$ or $\lambda \rightarrow 0$ at the depinning
transition (Fig. \ref{f:lambda}).  Thus, the scaling behavior of these
models should be correctly described by (\ref{qew}).  For the DPD-1,
DPD-2, and DPD-3 models we observe a divergent $\lambda$, indicating
that nonlinearities are relevant near the depinning transition.  Thus
to properly describe the scaling properties of these models it is
necessary to study (\ref{qkpz}), since (\ref{qew}) does not include
the nonlinear term $\lambda (\nabla h)^2$.  Further evidence of the
existence of the two universality classes is given by the values of
the roughness exponents.  The models for which $\lambda$ diverges at
the depinning transition, have $\alpha \approx 0.63$, in agreement with
the mapping to DP \cite{explain3}.  On the other hand, models in the
universality class of Eq. (\ref{qew}), gave roughness exponents
typically larger, and in better agreement with the theoretical
prediction, $\alpha \approx 1$.

  It is worthwhile to discuss the implications of the previous results
on other models and theoretical approaches studied in the literature.
Recently, considerable attention has been focused on a self-organized
version of the DPD model, the so-called self-organized depinning (SOD)
model \cite{Havlin,amaral2,Sneppen,sj,lt,olami,maslov}.  In this model,
the constant driving force of the DPD model is replaced by an algorithm
that selects the site on the interface for which the disorder is weakest
--- i.e, it selects the point in the interface offering the least
resistance.  Thus, the local update rule of the DPD model is replaced by
a global search process, similar to the one used in invasion
percolation.  This update rule assures that the model will evolve to
criticality and that it will remain in the critical state.  This implies
that the SOD model is always at the depinning transition.  Thus the
properties of the SOD model are described by the DPD model at
criticality, which implies, since the SOD model is always at
criticality, that the SOD model cannot be described by a continuum
equation of the KPZ-type, since this would require an infinite
coefficient of the nonlinear term.  However, this argument {\it does
not\/} rules out that the SOD model can be described by a continuum
equation that upon a mapping to the quenched KPZ equation will generate
an effective coefficient $\lambda$ that would diverge \cite{Tang95}.

\section{The Moving Phase}

\subsection{Why look at the moving phase?}

  In the previous section we identified the universality classes for
interface roughening in the presence of quenched disorder.  However,
the experimental situation still leaves us with several unsolved
puzzles, such as the change in the roughness exponent or the
dependence of the global width on the driving force for the
fluid-fluid displacement experiments.

  In trying to shed some light over the remaining problems, we observe
that most of the numerical and the analytical studies focus on the
``pinned phase'' ($F \le F_c$) while nearly all experimental results
are for the ``moving phase'' ($F > F_c$).  A logical next step is
therefore to study the moving phase for models representative of each
of the two universality classes.

\subsection{The DPD universality class}

\subsubsection{The steady state}

  Let us start by considering the DPD universality class.  In Fig.
\ref{sdpd}(a), we show the local width for the pinned and the moving
phases.  We find $\alpha_P \approx 0.63$ for the pinned phase, and
$\alpha_M \approx 0.75$ for the moving phase (Table \ref{tab_exp}).
So far, no explanation for this change in roughness exponent at the
depinning transition is available.

  The study of the scaling of the local width for the moving phase
reveals a {\it novel\/} dependence of the prefactor of the width on
the external force. We also find that the exponents characterizing
this anomalous dependence change values for the two different regimes.
In the first
\begin{equation}
w \sim \ell^ {\alpha_M}~f^{-\varphi_M} \qquad (\ell \ll \xi),
\label{mov1}
\end{equation}
where $\xi$ is the correlation length, and $\varphi_M$ is a new exponent
that characterizes the dependence of the prefactor of the width on the
driving force. In the second regime, the effect of the quenched disorder
becomes irrelevant compared to the annealed disorder \cite{expl0}, and
we obtain
\begin{equation}
w \sim \ell^ {\alpha_A}~f^{- \varphi_A} \qquad (\ell \gg \xi),
\label{mov2}
\end{equation}
where $\varphi_A$ is a new exponent, and $\alpha_A$ is the roughness
exponent corresponding to annealed disorder.  Depending on the absence
or presence of nonlinear terms, we recover the results of either the
EW or the KPZ equations with annealed disorder.  This crossover was
observed both for experiments and simulations of discrete models
\cite{alb}.

  Due to the dependence of the prefactor of the width on the driving
force, and since $\xi \sim f^{-\nu}$, we propose the scaling {\it
Ansatz}
\begin{equation}
w(\ell,f) \sim \ell^{\alpha_A}~ f^{- \varphi_A}~ g(\ell / \xi).
\label{scal_w}
\end{equation}
Upon comparison with (\ref{mov1}) and (\ref{mov2}), we find that the
scaling function $g(u)$ satisfies $g(u \gg 1) \sim ~$ const, and
$g(u~\ll~1) \sim u^{\alpha_M - \alpha_A}$. Using (\ref{scal_w}) we
also obtain
\begin{equation}
\varphi_A = \varphi_M + \nu (\alpha_M - \alpha_A).
\label{chi1}
\end{equation}
In Fig. \ref{sdpd}(b) we show the data collapse of the widths for
different forces and length scales.  The deviations from scaling for
large values of $\ell / \xi$ are due to finite-size effects (see
Fig. \ref{sdpd}(d) for discussion).

  While previous studies considered $\alpha_M$ as an effective
exponent, and concluded that no self-affine scaling exists in the
moving phase for $\ell \ll \xi$ \cite{Tang}, we find a consistent
value for $\alpha_M$ regardless of how closely we approach the
depinning transition.  The self-affinity of the interface is supported
by the good data collapse obtained with (\ref{scal_w}), as shown in
Fig. \ref{sdpd}.

  To derive a second relation for the new exponent $\varphi_M$, let us
consider the approach to the depinning transition for a system of size
$L$.  For a finite system, the transition does not occur for $f=0$, as
it would for an infinite system, but rather for an effective critical
force $F_c(L)$ such that $\xi \sim |F_c(L) - F_c|^{-\nu} \sim L$,
implying $f \sim L^{-1/ \nu}$.  Using this result and (\ref{mov1}), we
obtain
\begin{equation}
w \sim \ell^{\alpha_M}~L^{\varphi_M / \nu}  \qquad (\ell \ll L \sim \xi).
\label{mov}
\end{equation}
For the pinned phase we have
\begin{equation}
w \sim \ell^{\alpha_P} \qquad (\ell \ll L \sim \xi).
\label{pin}
\end{equation}
At the depinning transition, (\ref{mov}) and (\ref{pin}) should scale
in the same way.  If we replace $\ell$ by $L$ we obtain \cite{expl-scal}
\begin{equation}
\varphi_M = \nu (\alpha_P - \alpha_M).
\label{chi3}
\end{equation}
Replacing the known values of $\nu$, $\alpha_P$, $\alpha_M$, and
$\alpha_A$ into (\ref{chi1}) and (\ref{chi3}), we find
\begin{equation}
\varphi_M \approx -0.20, \quad \varphi_A \approx 0.23,
\end{equation}
to be compared with the measured values $\varphi_M \approx -0.12$ and
$\varphi_A \approx 0.34$ (Table \ref{tab_exp}).  Although the
agreement with the measured values is not perfect, the error bars do
not rule out the validity of (\ref{chi3}).

\subsubsection{The time evolution}

  For the DPD universality class, we find that $\beta$ changes value
at the depinning transition (see Table \ref{tab_exp}).  We find
$\beta_P \approx 0.63$ for the pinned phase and $\beta_M \approx 0.75$
for the moving phase.  We obtain for both sides of the transition
$\alpha \approx \beta$, implying that $z$ remains unchanged at
the transition \cite{beta}.  The exponent $z$ characterizes the time
scale $t_{\times}$ for the propagation of correlations in the
interface (cf. Eq. (\ref{scl1})). This time scale is not expected to
depend on the external force \cite{havlin2}.  These results are in
good agreement with a numerical integration of Eq. (\ref{qkpz})
\cite{csahok}.

  For the time evolution of the width, we again find an anomalous
dependence of the width on the reduced for the two different scaling
regimes. In regime I,
\begin{equation}
W \sim t^ {\beta_M}~f^{-\kappa_M} \qquad (t \ll t_{\times}),
\label{tmov1}
\end{equation}
where $t_{\times} \sim \xi^z$ is the characteristic time for the
propagation of correlations over a distance $\xi$, and $\kappa_M$ is a
new exponent that characterizes the dependence of the prefactor of the
width on the driving force. In regime II, the effect of the pinned
disorder becomes irrelevant compared to the annealed disorder
\cite{expl0}, and we obtain
\begin{equation}
W \sim t^ {\beta_A}~f^{- \kappa_A} \qquad (t \gg t_{\times}),
\label{tmov2}
\end{equation}
where $\kappa_A$ is a new exponent, and $\beta_A$ is the growth
exponent corresponding to annealed disorder.

  Since $t_{\times} \sim \xi^{z} \sim f^{- z \nu}$, we propose
the scaling {\it Ansatz}
\begin{equation}
W(t,f) \sim t^{\beta_A}~ f^{- \kappa_A}~ g(t / t_{\times}).
\label{scal_wt}
\end{equation}
Upon comparison with (\ref{tmov1}) and (\ref{tmov2}), we find that the
scaling function $g(u)$ satisfies $g(u \gg 1) \sim ~$ const, and $g(u
\ll 1) \sim u^{\beta_M - \beta_A}$.  From (\ref{scal_wt}) we also obtain
\begin{equation}
\kappa_A = \kappa_M + z \nu (\beta_M - \beta_A).
\label{kappa1}
\end{equation}
In Fig. \ref{tdpd} we display the rescaling of our simulation results
for the dependence of the global width on time and driving force
according to (\ref{scal_wt}).

  To determine a second relation for the new exponent $\kappa_M$, let
us again consider the approach to the depinning transition for a
system of size $L$.  As discussed above, the transition occurs for an
effective critical force such that $\xi \sim L$, implying $f \sim
L^{-1/ \nu} \sim t_{\times}^{-1 / z \nu}$.  Using this result and
(\ref{tmov1}), we obtain
\begin{equation}
W \sim t^{\beta_M}~t_{\times}^{\kappa_M / z \nu} \qquad
(t \ll t_{\times}).
\label{tmov}
\end{equation}
For the pinned phase, we have
\begin{equation}
W \sim t^{\beta_P} \qquad (t \ll t_{\times}).
\label{tpin}
\end{equation}
At the transition, (\ref{tmov}) and (\ref{tpin}) must scale in a
similar way. So, if we replace $t$ by $t_{\times}$, we obtain
\begin{equation}
\kappa_M = z \nu (\beta_P - \beta_M).
\label{kappa3}
\end{equation}
Replacing the known values of $\nu$, $z$, $\beta_P$, $\beta_M$, and
$\beta_A$ into (\ref{kappa1}) and (\ref{kappa3}), we find
\begin{equation}
\kappa_M \approx -0.12, \quad \kappa_A \approx 0.64,
\end{equation}
in good agreement with the simulation results, $\kappa_M \approx
-0.11$ and $\kappa_A \approx 0.65$ (Table \ref{tab_exp}).

\subsection{The QEW universality class}

\subsubsection{The steady state}

  For the study of the QEW universality class, we focus on the QEW-2
model.  We observe a dependence of the local width on the system size
$L$, that can be described by the scaling function \cite{lesc1,sarma}
\begin{equation}
w(\ell,L) \sim L^{\alpha_G} \Phi(\ell / L),
\label{self}
\end{equation}
where $\Phi(u)$ is a scaling function that for $u \ll 1$ scales as
$u^{\alpha_P}$, and $\alpha_G$ is the global roughness exponent, which
is defined by the scaling of the global width with the system size:
$W \sim L^{\alpha_G}$.  Equation (\ref{self}) is valid only when
$\alpha_G > 1$, and it implies that $w(\ell,L) \sim \ell^{\alpha_P}
L^{\alpha_G - \alpha_P}$ for $\ell \ll L$.  Hence, the local slopes
are unbounded, i.e., they diverge with the system size.  This
anomalous characteristic of the QEW models is displayed in
Fig. \ref{f:anom}(a), where we plot interfaces generated for the QEW-2
model for different values of the system size.  We calculate the
global width as a function of $L$ and find at the depinning transition
that $\alpha_G \approx 1.23$.

  We also determine the scaling of the local width $w$ in a window of
size $\ell$, for different values of the reduced force.  In
Fig. \ref{sqew}(a), we show the local width for the pinned and the
moving phases.  Analysis of consecutive slopes yield roughness
exponents $\alpha_P \approx 0.92$ for the pinned phase, and $\alpha_M
\approx 0.92$ for the moving phase (Table \ref{tab_exp}).  These
results are in agreement with values commonly found for the QEW class
of models when the roughness exponent is calculated from the local
width --- i.e., $\alpha_P \approx \alpha_M \approx 1$.  The slightly
smaller value of our estimates is is likely due to finite size
effects.

  As for the DPD universality class, the study of the local width in
the moving phase reveals an anomalous dependence on the reduced force,
which leads to Eqs. (\ref{scal_w}) and (\ref{chi1}) being also valid
for the QEW universality class.  In Fig.  \ref{sqew}(b), we show the
data collapse of the local widths obtained using (\ref{scal_w}).

  To determine a second relation for the new exponent $\varphi_M$, we
again consider the approach to the depinning transition of a system of
size $L$.  As before, we obtain (\ref{mov}) for the moving phase.
However, for the pinned phase, the scaling behavior (\ref{self}),
leads to
\begin{equation}
w \sim \ell^{\alpha_P}~L^{\alpha_G - \alpha_P}
\qquad (\ell \ll L).
\label{pin0}
\end{equation}
At the transition, (\ref{mov}) and (\ref{pin0}) must be identical.
Hence, we find
\begin{mathletters}
\begin{equation}
\alpha_P = \alpha_M,
\end{equation}
and
\begin{equation}
\varphi_M = \nu (\alpha_G - \alpha_P).
\end{equation}
\label{chi2}
\end{mathletters}
Replacing the measured values of $\nu$, $\alpha_G$, $\alpha_P$, and
$\alpha_A$ into (\ref{chi1}) and (\ref{chi2}), we find
\begin{equation}
\varphi_M \approx 0.42, \quad \varphi_A \approx 1.04,
\end{equation}
in good agreement with the values obtained in our simulations,
$\varphi_M \approx 0.44$ and $\varphi_A \approx 0.99$ (Table
\ref{tab_exp}).

\subsubsection{The time evolution}

  We study the scaling of $W$ with $t$, and find $\beta_P \approx
0.85$ and $\beta_M \approx 0.86$.  These results imply that $z$
remains unchanged at the transition.  In Fig.  \ref{tqew}, we show the
scaling of the global width as a function of time for different values
of the reduced force, and its data collapse using (\ref{scal_wt}).

  For the QEW universality class, we find by replacing the measured
values of $\nu$, $z$, $\beta_P$, and $\beta_A$ into (\ref{kappa1}) and
(\ref{kappa3}) that
\begin{equation}
\kappa_M \approx 0.02, \quad \kappa_A \approx 1.19,
\end{equation}
in good agreement with the values obtained in our simulations,
$\kappa_M \approx 0$ and $\kappa_A \approx 1.15$ (Table \ref{tab_exp}).

\subsection{Discussion}

  The results obtained in the previous subsections allow us to discuss
the experimental and numerical results reported in the literature.  A
problem with the interpretation of experimental and numerical results
for the roughness exponent, has been the wide range of values for
$\alpha$: $0.5-1.25$, measured in the moving phase.  We note that, in
this regime, the crossover to the annealed disorder regime leads to
``effective'' exponents that change with the velocity (or the driving
force).  For this reason, we suggest that the scaling function
(\ref{scal_w}) might be useful in the determination of the exponents
from the study of the local width $w$.  As shown in Table
\ref{tab_exp}, the exponent $\varphi_M$ has different signs for the two
universality classes, leading to sharply distinct scaling behaviors
for the prefactor of the width with the driving force.  Since in many
experiments it is possible to monitor the velocity of the interface,
and therefore the driving force, the study of this prefactor may also
lead to an easier identification of the universality class to which
the experiments belong.

  In light of this discussion, the interpretation of the results of
Ref.  \cite{kessler} is clear.  The numerical integration of the EW
equation with quenched disorder performed in Ref. \cite{kessler} must
belong to the QEW universality class, and the reported exponents are
``effective'' exponents whose values were affected by the crossover to
the annealed regime.

 The determination of the universality class of the {\it fluid
invasion model\/} (FIM) of Ref. \cite{robbins1,robbins2} is not
trivial.  However, we note that $\nu$ is identical for the FIM and for
the Hamiltonian model.  On the other hand, the value of $\alpha$
measured in Ref. \cite{robbins1,robbins2} is somewhat smaller than the
value obtained for the QEW model using the local width.

  Regarding the fluid-fluid displacement experiments of Refs.
\cite{stokes,rubio,horv,wong}, we find that they share several scaling
properties with the Hamiltonian model.  We first notice that the range
of roughness exponents measured in the experiments of
Refs. \cite{stokes,rubio,horv,wong} is consistent with the values that
could be obtained with the Hamiltonian model.  Moreover, Rubio {\it et
al.}  \cite{stokes,rubio,horv,wong} found that the amplitude of the
width decreases with the increase of the velocity. This property is
entirely consistent with our findings for the QEW universality class:
if we replace $f \sim v^{1/\theta}$ in (\ref{mov1}), then we obtain
that the width decreases with $v$ as $w \sim v^{-\varphi_M / \theta}$.

  The imbibition experiments of Refs. \cite{Havlin,Buldyrev,amaral1}
and the paper burning experiments of Ref. \cite{zzhang} are believed
to belong to the DPD universality class.  This conclusion is supported
by the values of the exponents measured in both experiments.

  The reason for the different universality classes for the two groups
of experiments, fluid-fluid displacement on one hand and imbibition
and paper burning on the other, is not totally clear.  However, a
possible explanation might be the importance of lateral growth in the
second case compared with the first.  When we analyze the imbibition
and paper burning experiments, it becomes clear that the interface can
be described by a single-valued interface, because only the highest
position of the interface in each column is relevant for the dynamics.
This effectively corresponds to eroding any overhangs and, as shown by
Makse \cite{makse3}, leads to a diverging $\lambda$ at the transition.
On the other hand, in the fluid-fluid displacement experiments, we
must at all times consider the full interface, because of the
trapped fluid that might exist below the interface.  Since it will
require some effort to displace this fluid or to separate it from the
rest of the displaced fluid, we cannot physically justify a rule such
as the erosion of overhangs. This implies that $\lambda$ vanishes at
the depinning transition, and the experiments must belong to the QEW
universality class.

\section{Summary}

  To summarize, we perform a systematic study of several models
proposed to understand the motion of driven interfaces in disordered
media.  We are able to identify {\it two\/} distinct universality
classes.  For one of these universality classes, QEW, we observe that
it can be described by a Langevin equation similar to the EW equation
but with quenched disorder.  For the QEW universality class nearly all
exponents for the pinned phase can be obtained by renormalization
group calculations.

  For the other universality class, DPD, we find that it can be
described by a Langevin equation similar to the KPZ equation, but with
quenched disorder.  Furthermore, we find that the coefficient of the
nonlinear term $\lambda$ diverges at the depinning transition.
Because of this divergence, no perturbative analytical calculations
have been possible so far.  However, a mapping of the static
properties of the interface, at the depinning transition, to DP yields
all static exponents of the problem.  A mapping of the dynamics to
isotropic percolation yields the dynamical exponent.

  The identification of these two universality classes leads to some
understanding of many of the results obtained previously for models
and experiments.  However, many other results could not be understood
in terms of the exponents obtained for the two universality classes
{\it at\/} the depinning transition.  For this reason, we also
investigated the scaling properties of the models in the two
universality classes {\it above\/} the depinning transition.

  We find that for the DPD universality class, $\alpha$ and $\beta$
change their values at the depinning transition. Furthermore, we find
a dependence of the prefactor of the width on the driving force.  The
new exponent $\varphi_M$, characterizing the scaling of this
prefactor, can be used to relate the values of the roughness exponents
on both sides of the transition.

  For the QEW universality class, we find that $\alpha$ and $\beta$
remain unchanged at the depinning transition.  As for the DPD
universality class, we also find a dependence of the prefactor of the
width on the driving force.  The exponent $\varphi_M$ is, in this
case, found to relate the different values of the local, $\alpha_P$,
and global, $\alpha_G$, roughness exponents at the depinning
transition.

  These results provide us with a more consistent understanding of the
scaling properties of the two universality classes both at and above
the depinning transition.  This knowledge enables us to interpret most
of the experimental results obtained previously.  However, several
questions, that demand an answer, are still unsolved.  One of them is
the reason for the change in values of the roughness exponent at the
depinning transition for the DPD universality class.  Another is the
understanding of the different values of $\alpha_G$ and $\alpha_P$ for
the QEW universality class.

\acknowledgments

  We thank S.~V. Buldyrev, R. Cuerno, S.~T. Harrington, S. Havlin, V.~K.
Horv\'ath, P.~Ch. Ivanov, K.~B. Lauritsen, H. Leschhorn, I. Procaccia,
P. Rey, R. Sadr, S. Tomassone, T. Vicsek, and P.-z. Wong for valuable
contributions and discussions.  L.A.N.A. acknowledges support from Junta
Nacional de Investiga\c c\~ao Cient\'{\i}fica e Tecnol\'ogica.  The
Center for Polymer Studies is supported by the National Science
Foundation.

\appendix
\section{Predicting the Exponents of the DPD Model}

\subsection{Directed percolation}

  Near $p_c$, the size of DP clusters is characterized by a
longitudinal (parallel) correlation length $\xi_{\parallel}$ and a
transverse (perpendicular) correlation length $\xi_{\perp}$ that
diverge as \cite{perc1,perc2}
\begin{equation}
\xi_{\parallel} \sim |p_c-p|^{-\nu_{\parallel}}, \qquad \xi_{\perp} \sim
|p_c-p|^{-\nu_{\perp}}.
\label{xi}
\end{equation}
The parallel and perpendicular correlation length exponents for DP
clusters have been calculated \cite{EE}, with the results
\begin{equation}
\nu_{\parallel} = 1.733 \pm 0.001, \quad \nu_{\perp} = 1.097 \pm 0.001
\qquad (d = 1).
\label{exp0}
\end{equation}

\subsection{Static properties}

  The mapping of the {\it pinned interface\/} to DP enables us to
estimate the static exponents of this problem from the characteristic
exponents of DP clusters.  The characteristic length $\xi$ of the
pinned regions must be of the order of $\xi_{\parallel}$, so we can
identify the exponent $\nu$ to be
\begin{equation}
\nu = \nu_{\parallel}.
\label{nu}
\end{equation}

  The global width $W_{\rm sat}$ of the pinned interface should scale
as $\xi_{\perp}$, since its advance is blocked by a DP path.  On the
other hand, $\xi_{\parallel}$ must be larger than the system size $L$
for the interface to become pinned, from which follows
\cite{Havlin,Buldyrev,Tang}
\begin{equation}
W_{\rm sat} \sim \xi_{\perp} \sim \xi_{\parallel}^{\nu_{\perp} /
\nu_{\parallel}} \sim L^{\nu_{\perp} / \nu_{\parallel}} \qquad
(\xi_{\parallel} \ge L).
\label{eq100}
\end{equation}
Comparing with (\ref{wsat}), we conclude that the roughness exponent
is given in terms of the correlation exponents for DP,
\begin{equation}
\alpha = \nu_{\perp}/\nu_{\parallel}.
\label{alp}
\end{equation}
Substituting (\ref{exp0}) into (\ref{alp}), we predict
\begin{equation}
\alpha = 0.633 \pm 0.001 \qquad (d = 1).
\label{alpxx}
\end{equation}

\subsection{The minimum path}

  For any fractal set with a fractal dimension smaller than the
dimension of the embedding space, the distance between to points of
the set $\ell$ is not given by the Euclidean distance $r$ between
those points but by the minimum path distance (or chemical distance)
$\ell_{\rm min}$.  Numerical studies show that $\ell_{\rm min}$ scales
as \cite{perc1,perc2}
\begin{equation}
\ell_{\rm min} \sim r^{d_{\rm min}},
\end{equation}
where $d_{\rm min}$ is called the {\it minimum path\/} exponent.

\subsection{Dynamical properties}

  Recently, Havlin {\it et al.} showed that the dynamics of the DPD
model in $d+1$ dimensions could be mapped to the minimum path of
isotropic percolation in $d$ dimensions \cite{havlin2}.  They argued that
close to the depinning transition only a few cells on the interface are
active, i.e., unblocked.  These active cells perform a global search
of unblocked cells and propagate correlations in the interface.  Since
the cells through which the active cells can move are confined to
unblocked or eroded blocked cells in a region of thickness
$\xi_{\perp} \ll \xi_{\parallel}$, we can identify this region with an
isotropic percolation cluster embedded in a space of dimension $d$.

  Thus, the path through which active cells move is the minimum path.
So, the time $t$ for the correlations to propagate an Euclidean
distance $r$, is proportional to $\ell_{\rm min}$, which scales as
$r^{d_{\rm min}}$ for isotropic percolation.  From this follows that
we can connect the dynamical exponent $z$ to the static exponent
$d_{\rm min}$ of {\it isotropic\/} percolation \cite{havlin2}
\begin{equation}
z = d_{\rm min}.
\end{equation}

\section{ The DPD-3 Model }

  In this appendix, we discuss the model introduced by Parisi to study
interface roughening in the presence of quenched disorder
\cite{parisi}.  We show that the original growth rule leads to an
unphysical behavior.  We then proceed to describe the model proposed by
Amaral that solves the problems with the original model \cite{amaral3}.

  Let us start by describing the model originally introduced by
Parisi.  In a square lattice of edge $L$, with periodic boundary
conditions in that direction, we assign to each cell an uncorrelated
random number, uniformly distributed in the interval $[-1,1]$.  We
start, at $t = 0$, by invading the bottom row in the lattice, and
then, we calculate the driving force for each column
\begin{equation}
F(i,t) = \rho~{\rm max}[h(i\pm1,t) + 1 - h(i,t)],
\label{force}
\end{equation}
where $\rho$ is the controlling parameter of the model.  If this
driving force $F_i$ is larger than the noise in the cell above the
interface, $\eta(i,h(i)+1)$, then we update the height of column $i$ to
\begin{equation}
h(i,t+1) = {\rm max}[h(i\pm1,t) + 1].
\label{grow1}
\end{equation}

  The time evolution of this model leads to the development of huge
jumps in the height of neighboring columns that propagate along the
system.  For small enough values of $\rho$, only one or two such
growing regions will be propagating in the system.  The reason for
such ``unnatural'' behavior of the model is the growth rule
(\ref{grow1}).  The fact that the advancing column grows to an higher
value than the highest neighbor is a kind of ``bootstrap'' that is
difficult to justify.  In fact, it does not seem reasonable that any
growth mechanism would pull a lower column to an height larger than
that of its neighbors.

  For this reason, Amaral proposed a new growth rule to replace
(\ref{grow1})
\begin{equation}
h(i,t+1) = h(i,t) + 1.
\label{grow2}
\end{equation}
In this way, the invading region only advances one cell at a time;
thus, avoiding the formation of big jumps between neighboring
columns \cite{amaral3}.

  The study of this model, which we will refer to as ``DPD-3'', leads to
the result that a critical value of the controlling parameter exists,
$\rho_c = 0.196 \pm 0.002$.  Below this threshold the interface stops
moving after some finite time, and above it it moves indefinitely.  To
determine some of the scaling exponents, we applied the gradient method,
introduced by Sapoval {\it et al.} \cite{sapoval}, and already used in
the study of the DPD-2 model \cite{amaral1,amaral2} and in several other
studies \cite{wilk,birol,hansen}.

  Figure \ref{grad} displays our results for the case of a linear
gradient, $\nabla \rho (h) = g$.  The scaling of the interface changes
with the concentration rate and is characterized by an exponent
$\gamma$ according to
\begin{equation}
\label{scaling}
w(\ell,g) \sim \ell^\alpha f \left ( \ell / g^{-\gamma / \alpha}
\right )
\end{equation}
where $f$ is an universal scaling function that satisfies $f(x\ll
1) \sim $ const, and $f(x \gg 1) \sim x^{-\alpha}$.  Figure
\ref{grad}(a) shows the width $w$ as a function of the size $\ell$ for
different values of the concentration gradient $g$. Figure
\ref{grad}(b) shows the data collapse of these results according to
(\ref{scaling}).  The best collapse was obtained with the exponents
\begin{equation}
\alpha = 0.60 \pm 0.05, \quad \gamma = 0.49 \pm 0.05.
\label{exp1}
\end{equation}
Since we know that \cite{amaral1}
\begin{equation}
\alpha = \nu_{\perp} / \nu_{\parallel}, \quad
\gamma = \nu_{\perp} / ( 1 + \nu_{\perp} ),
\end{equation}
we obtain
\begin{equation}
\nu_{\perp} = 1.0 \pm 0.1, \quad \nu_{\parallel} = 1.6 \pm 0.1,
\end{equation}
in agreement with the known results for DP \cite{EE}.

\begin{figure}
\narrowtext
\caption{  The depinning transition.  In the ``pinned phase,'' $F <
F_c$, the velocity of the interface is zero.  In the ``moving phase,''
$F > F_c$, the interface moves with an average velocity $v \equiv
v(F)$, where $v(F) \sim (F - F_c)^{\theta}$ for $F$ close to $F_c$, and
$v(f) \sim F$ for $F \gg F_c$.  Thus, the velocity plays the role of
the order parameter of the transition.  }
\label{phases}
\end{figure}

\begin{figure}
\narrowtext
\caption{  Values of the roughness exponent $\alpha$ from most of the
experiments,simulations and theoretical works reported in the
literature.  The experimental points are, reading from left to right,
from Refs. [8], [9], [10] (a range of values is reported of which the
extremes are plotted), [12-17], [11].  The simulation points are,
reading from left to right, from Refs. [17] (a range of values is
reported of which the extremes are plotted), [29], [12-17], [32],
[31], [34], [33], [40], [27-28], [45], [46].  Finally, the theoretical
points, again reading from left to right, are from Refs. [48] and
[49].  Visually apparent is the wide spread of the results.  To guide
the eye, we indicate the values of $\alpha$ predicted by the KPZ
equation with annealed disorder (bottom line), the DPD universality
class in the pinned phase and moving phases (middle lines), and the
QEW universality class (top line).  We can see that most results for
simulations and theoretical calculations are close to either the DPD
or the QEW line.  However, for the experimental results, the agreement
is not so good.  In the text we argue that the reasons for these
disagreements are the crossover present in the moving phase for both
universality classes, and the fact that for the DPD universality class
$\alpha_M \ne \alpha_P$.  }
\label{f:summary}
\end{figure}

\begin{figure}
\caption{  {\it DPD-1 model\/}. (a) Dependence on the tilt $m$ of the
average velocity in $1+1$ dimensions.  Data for different values of
the force $f = (F - F_c) / F_c$ are indicated by different symbols,
ranging from $0.016$ (bottom curve) to $0.350$ (top curve).  The
system size is $L = 512$ and each result was averaged over $30$
realizations of the disorder.  (b) Data collapse of the velocities
according to (4.7) using the values for the DPD-3 model exponents
(Table I). }
\label{dpd1}
\end{figure}

\begin{figure}
\caption{  Here we exemplify the ``noncrossing'' effect on the velocity
parabolas. We show a perfect parabolic behavior for two different
forces, $f_1>f_2$ (dashed lines) as predicted by Eq. (4.3).  Also
shown is the ``curving back'' of the velocity curve for the smaller
force $f_2$ (solid line) which is necessary in order not to cross the
velocity curve for $f_1$.  }
\label{cross}
\end{figure}

\begin{figure}
\caption{  {\it DPD-2 model\/}. (a) Dependence on the tilt $m$ of the
average velocity in $1+1$ dimensions.  Data for different forces $f$
are indicated by different symbols, ranging from $0.0149$ (bottom
curve) to $0.0719$ (top curve).  The system size is $512$ and each
result was averaged over $30$ realizations of the disorder.  (b) Data
collapse of the velocities according to (4.7) using the values for the
DPD-2 model exponents (Table I). }
\label{dpd2}
\end{figure}

\begin{figure}
\caption{  {\it DPD-3 model\/}. (a) Dependence on the tilt $m$ of the
average velocity in $1+1$ dimensions.  Data for different values of
the force $f$ are indicated by different symbols, ranging from
$0.0204$ (bottom curve) to $0.2454$ (top curve).  The system size is
$L = 1024$ and each result was averaged over $30$ realizations of the
disorder.  (b) Data collapse of the velocities according to (4.7)
using the values for the DPD-3 model exponents (Table I). }
\label{dpd3}
\end{figure}

\begin{figure}
\caption{  {\it QEW-1 model\/}. Dependence on the tilt $m$ of the
average velocity in $1+1$ dimensions.  Data for different values of
the force $f$ are indicated by different symbols, ranging from
$0.0125$ (bottom curve) to $0.075$ (top curve).  The system size is $L
= 1024$ and each result was averaged over $30$ realizations of the
disorder.  We can see clearly that $\lambda = 0$.  }
\label{qew1}
\end{figure}

\begin{figure}
\caption{  {\it RFIM\/}. (a) Dependence on the tilt $m$ of the average
velocity in $2+1$ dimensions.  Data for different values of the force
$f$ are indicated by different symbols, ranging from $0.014$ (bottom
curve) to $0.143$ (top curve).  The system size is $L^2 = 40 \times
40$, and $\Delta = 3$ (SA regime).  (b) Data collapse of the
velocities according to (4.7) using the values for the RFIM exponents
(Table I).  }
\label{rfim2}
\end{figure}

\begin{figure}
\caption{  Schematic representation of the scaling behavior of the
coefficient $\lambda$ of the nonlinear term for the two universality
classes.  For the DPD universality class we see that $\lambda$
diverges at the depinning transition, while for the QEW universality
class it vanishes to zero at the transition.  }
\label{f:lambda}
\end{figure}

\begin{figure}
\narrowtext
\caption{  {\it DPD universality class}.  (a) Plot of the local width
$w$ as a function of $\ell$ for a pinned and a moving interface.  The
different roughness exponent is clear.  The system size is $L = 10480$
and each result was averaged over $50$ realizations of the disorder.
(b) Plot of the local width in the moving phase for several values of
the driving force.  (c) Data collapse of the widths according to (5.3).
(d) Data collapse for $f = 0.0583$, and for systems of different size
$L$.  Visually apparent is the fact that the deviations from scaling
observed in (c) are due to finite size effects.  }
\label{sdpd}
\end{figure}

\begin{figure}
\narrowtext
\caption{  {\it DPD universality class}.  (a) Plot of the global width
$W$ as a function of time the moving phase interface.  The system size
is $L = 10480$ and each result was averaged over $50$ realizations of
the disorder.  (b) Data collapse of the widths according to (5.11).  }
\label{tdpd}
\end{figure}

\begin{figure}
\narrowtext
\caption{  Anomalous scaling of the local width for the QEW universality
class.  (a) Interfaces generated with the QEW-2 model, at the
depinning transition, for system of sizes $L =128, 256, 512, 1024$.
We note that the local slopes increase with the system size.  (b) Plot
of the scaling of the local width as a function of $L$, for two given
values of $\ell$.  The fit corresponds to $\alpha_G - \alpha_P \approx
0.25$.  }
\label{f:anom}
\end{figure}

\begin{figure}
\narrowtext
\caption{  {\it QEW universality class}.  (a) Plot of the local width
in the moving phase for several values of the reduced force.  The system
size is $L = 5524$ and each result was averaged over $100$ realizations
of the disorder.  (b) Data collapse of the widths according to (5.3).  }
\label{sqew}
\end{figure}

\begin{figure}
\narrowtext
\caption{  {\it QEW universality class}.  (a) Plot of the global width
$W$ as a function of time for the moving phase.  The system size is $L =
5524$ and each result was averaged over $100$ realizations of the
disorder.  (b) Data collapse of the widths according to (5.11).  }
\label{tqew}
\end{figure}

\begin{figure}
\narrowtext
\caption{  Simulation results for the width $w(\ell,g)$ of the
pinned interface in $1+1$ dimensions, where $g$ is the gradient in
$\rho$.  (a) The widths for values of the gradient ranging from $1/70$
to $1/7680$.  The system size is $10480$ and each result is averaged
over 30 realizations of the disorder.  (b) Data collapse of these
results according to the scaling form (B4), using the values of the
exponents given in (B5).  }
\label{grad}
\end{figure}

\end{multicols}

\begin{table}
\caption{  Exponents for the six models studied (see definitions in
the text).  A negative value of $\phi$ means that $\lambda \rightarrow
0$ when $f \rightarrow 0$ [cf. Eq. (4.5)].  We argue in the text that
the models above the horizontal line (DPD-1, DPD-2, and DPD-3) belong to
the universality class of Eq. (3.3) and can be mapped, in $1+1$
dimensions, to DP.  The models below the line belong to the universality
class of Eq. (3.1).  See [56] for a discussion on the values of the
exponents for the DPD-2 model.}
\begin{tabular}{llcccc}
\tableline
\multicolumn{2}{l}{Model} & \multicolumn{2}{c}{$1+1$ dimensions} &
  \multicolumn{2}{c}{$2+1$ dimensions} \\ & & $\theta$ & $\phi$ &
$\theta$ & $\phi$ \\
\tableline
\tableline
DPD-1    &       & $0.64\pm0.08$ & $0.64\pm0.08$ & $0.80\pm0.12$ &
$0.30\pm0.12$   \\
DPD-2    &       & $0.59\pm0.12$ & $0.55\pm0.12$ &       &       \\
DPD-3    &       & $0.70\pm0.12$ & $0.65\pm0.12$ &       &       \\
\tableline
QEW-1   &       & $0.26\pm0.07$ & ---           &       &       \\
QEW-2   &       & $0.24\pm0.03$ & ---           &       &       \\
        & FA    & $0.9\pm0.2$   & $1.8\pm0.2$   &       &       \\
RFIM    & SA    & ---   & ---   & $0.60\pm0.11$ & $-0.70\pm0.11$
\\
        & SS    & $0.31\pm0.08$ & $-0.65\pm0.13$&       &       \\
\tableline
\end{tabular}
\label{tab1}
\end{table}

\begin{table}
\narrowtext
\caption{ Critical exponents for the two universality classes studied
in this paper.  The subscript ``$P$'' refers to the {\it pinned\/}
phase, the subscript ``$M$'' denotes the quenched disorder regime in
the {\it moving\/} phase, and the subscript ``$A$'' refers to the {\it
annealed disorder\/} regime.  The exponents $z$ and $z_A$ were
calculated from scaling relations, while the remaining exponents were
calculated directly in the simulations.  }
\begin{tabular}{lcc}
Exponents        & DPD                   & QEW   \\
\tableline
$~\alpha_P$      & $0.63\pm0.03$         & $0.92\pm0.04$ \\
$~\alpha_M$      & $0.75\pm0.04$         & $0.92\pm0.04$ \\
$~\alpha_A$      & $0.50\pm0.04$         & $0.46\pm0.04$ \\
$~\alpha_G$      & ---                   & $1.23\pm0.04$ \\
\tableline
$~\beta_P$       & $0.67\pm0.05$         & $0.85\pm0.03$ \\
$~\beta_M$       & $0.74\pm0.06$         & $0.86\pm0.03$ \\
$~\beta_A$       & $0.30\pm0.04$         & $0.25\pm0.04$ \\
\tableline
$~z$             & $1.01\pm0.10$         & $1.45\pm0.07$ \\
$~z_A$           & $1.67\pm0.26$         & $2.09\pm0.40$ \\
\tableline
$~\nu$           & $1.73\pm0.04$         & $1.35\pm0.04$ \\
$~\theta$        & $0.64\pm0.12$         & $0.24\pm0.03$ \\
\tableline
$~\varphi_M$     & $-0.12\pm0.06~~$      & $0.44\pm0.05$ \\
$~\varphi_A$     & $0.34\pm0.06$         & $0.99\pm0.05$ \\
\tableline
$~\kappa_M$      & $-0.11\pm0.06~~$      & $0.00\pm0.06$ \\
$~\kappa_A$      & $0.65.\pm0.06$        & $1.15\pm0.06$ \\
\end{tabular}
\label{tab_exp}
\end{table}


\begin{references}
\bibitem{expl0}
We introduce this time-dependent randomness in the model to mimic the
experimental fact that for $F > F_c$, a transition to the KPZ behavior
with annealed disorder is observed for $\ell \gg \xi$.  In fact, Eqs.
(3.1) and (3.3) should also include annealed disorder terms.  However,
this annealed disorder is much weaker than the quenched disorder.  For
this reason, for $\ell \ll \xi$ the presence of time
dependent disorder is irrelevant.  A different situation occurs for
$\ell \gg \xi$. In this case the effect of quenched disorder is no
longer relevant, and even a week annealed disorder becomes important.
\bibitem{explain00}
The consistency of the argument developed in the text can be verified
by expanding (4.8) to first order, whereupon we recover (4.6).
\bibitem{explain0}
The justification of this particular rule lies not only on the
computational simplifications that it affords but also on the
experimental finding that overhangs do not play an important role in
the scaling properties of the interface for imbibition.
\bibitem{explain1}
As was shown in Refs. \cite{Havlin,Buldyrev,amaral2,Tang}, the DPD-1
and DPD-2 models belong to the same universality class, with exponents
$\alpha \approx 0.63$ and $\theta \approx 0.64$ \cite{amaral4}.  The
deviation of the numerical values in the exponents for the DPD-2 model
can be explained by the slow
convergence of this model to the asymptotic scaling regime.  The large
error bars for all models arise from the fact that simulations were
done for relatively small systems.  Despite the differences in the
values of the scaling exponents, based on the numerical results
combined with the exact results of Refs.
\cite{Havlin,Buldyrev,amaral2,Tang}, we conclude that
the models, DPD-1, DPD-2, and DPD-3, share the same set of scaling
exponents, thus belonging to the same universality class.
\bibitem{expl-fa}
Our results show that for the FA regime, the RFIM behaves in a similar
fashion to the models in the DPD universality class, in that the
coefficient of the nonlinear term diverges at the depinning
transition.  However, we stress that this regime has little physical
motivation, being its existence due to lattice effects.
\bibitem{expl-scal}
Since we want to calculate the scaling exponents we only need to
consider the leading order terms.  Hence, although Eqs. (\ref{mov})
and (\ref{pin}) are only valid for $\ell \ll L$, we can, to leading
order, compare the two expressions when $\ell = L$.
\bibitem{beta}
Reference \cite{Tang} measured $\beta_M \approx 0.63$. However,
the large intrinsic width \cite{kertesz} of their variant of the
DPD model, leads the authors of Ref. \cite{Tang} to add a constant
to their data, which probably altered the scaling of $W$ for
$t \ll t_{\times}$.
\end{references}
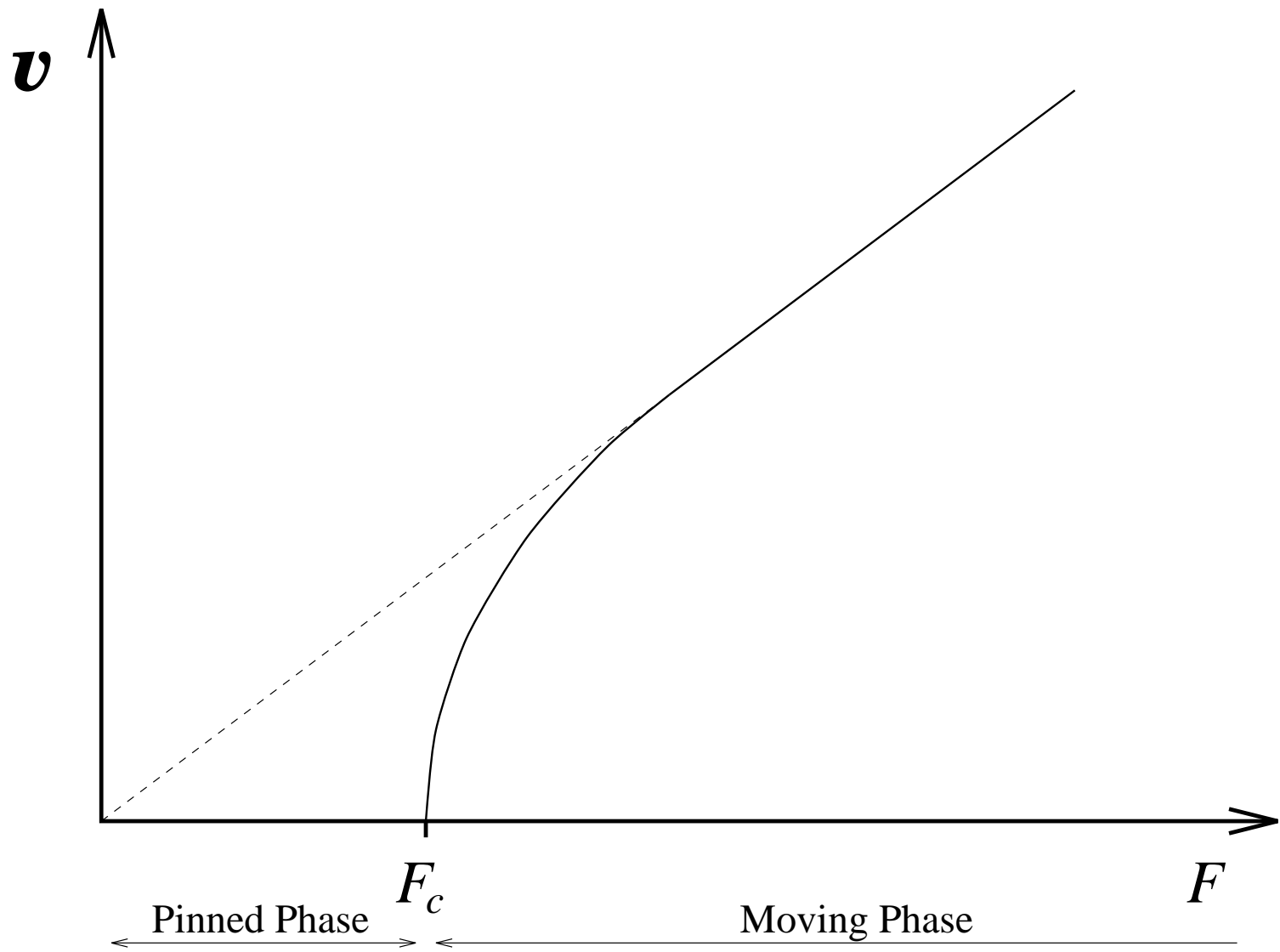
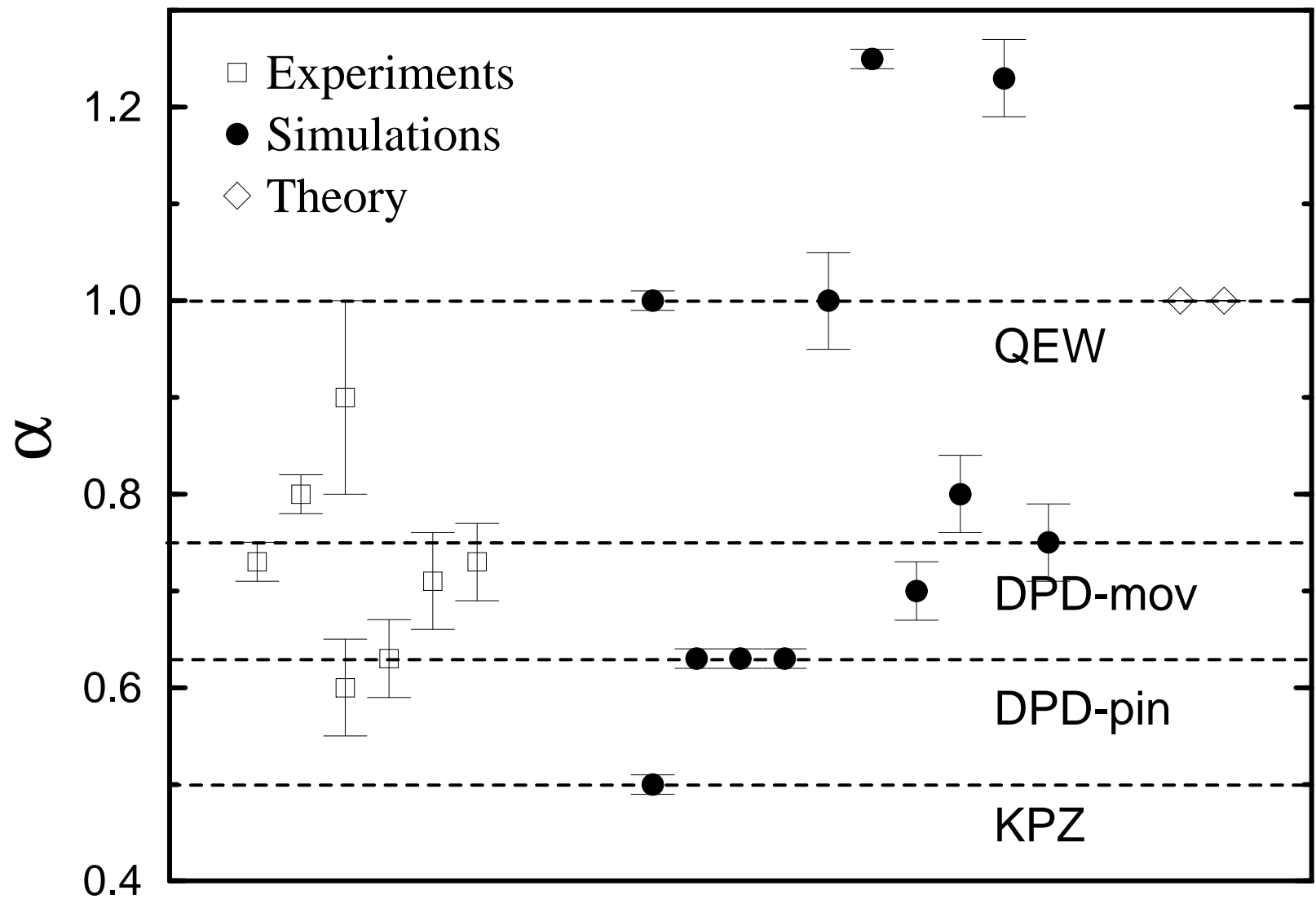
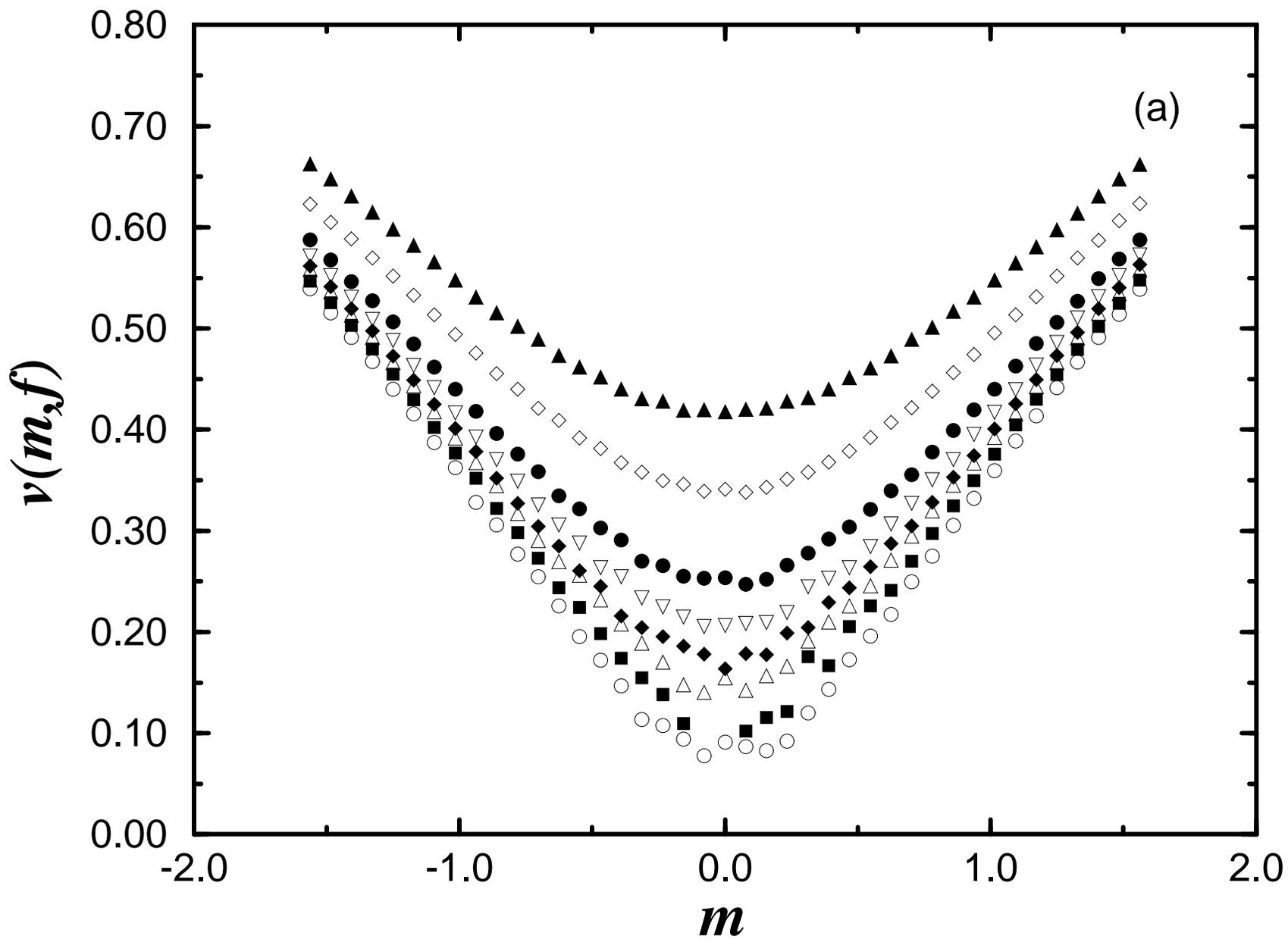
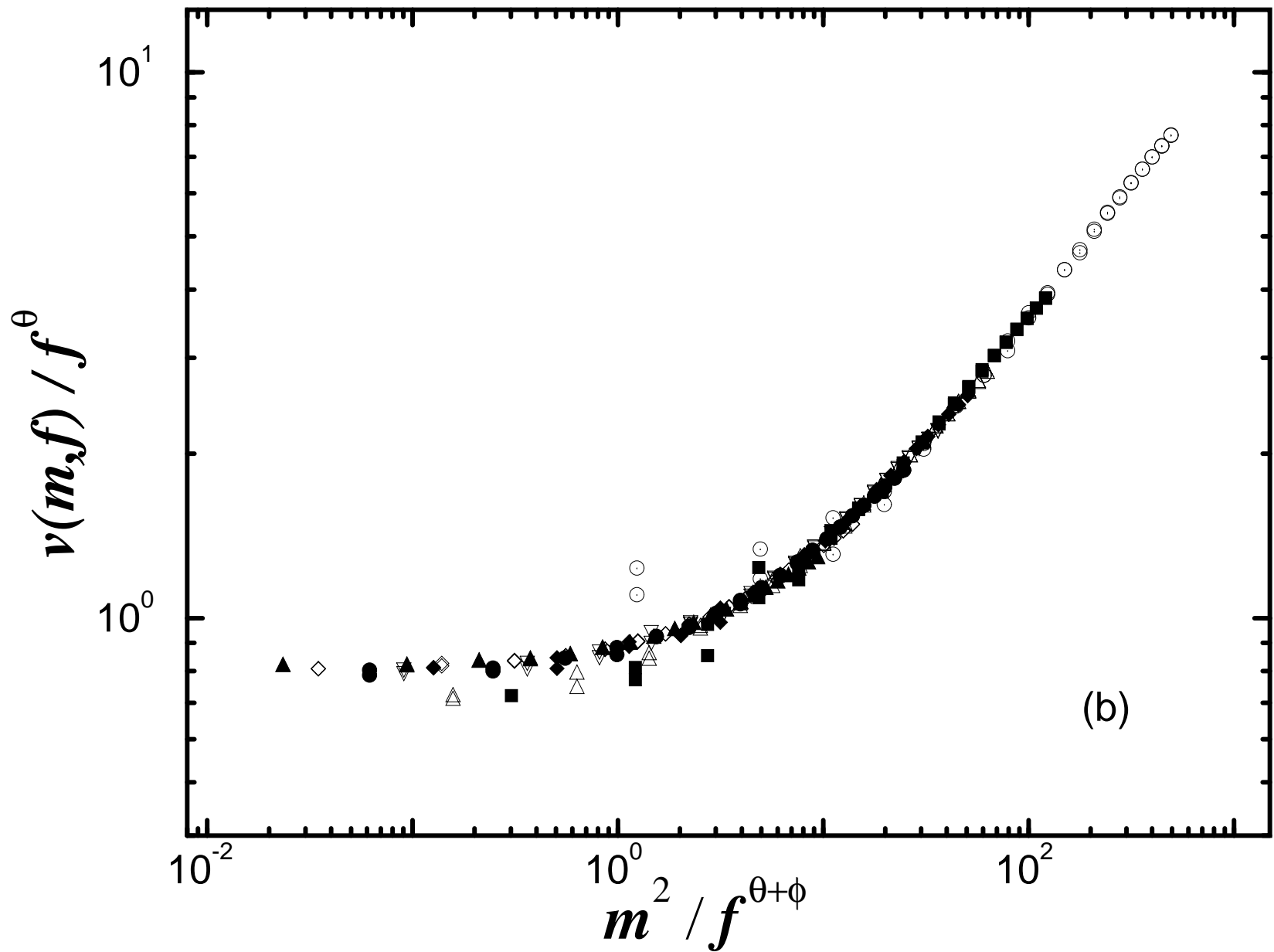
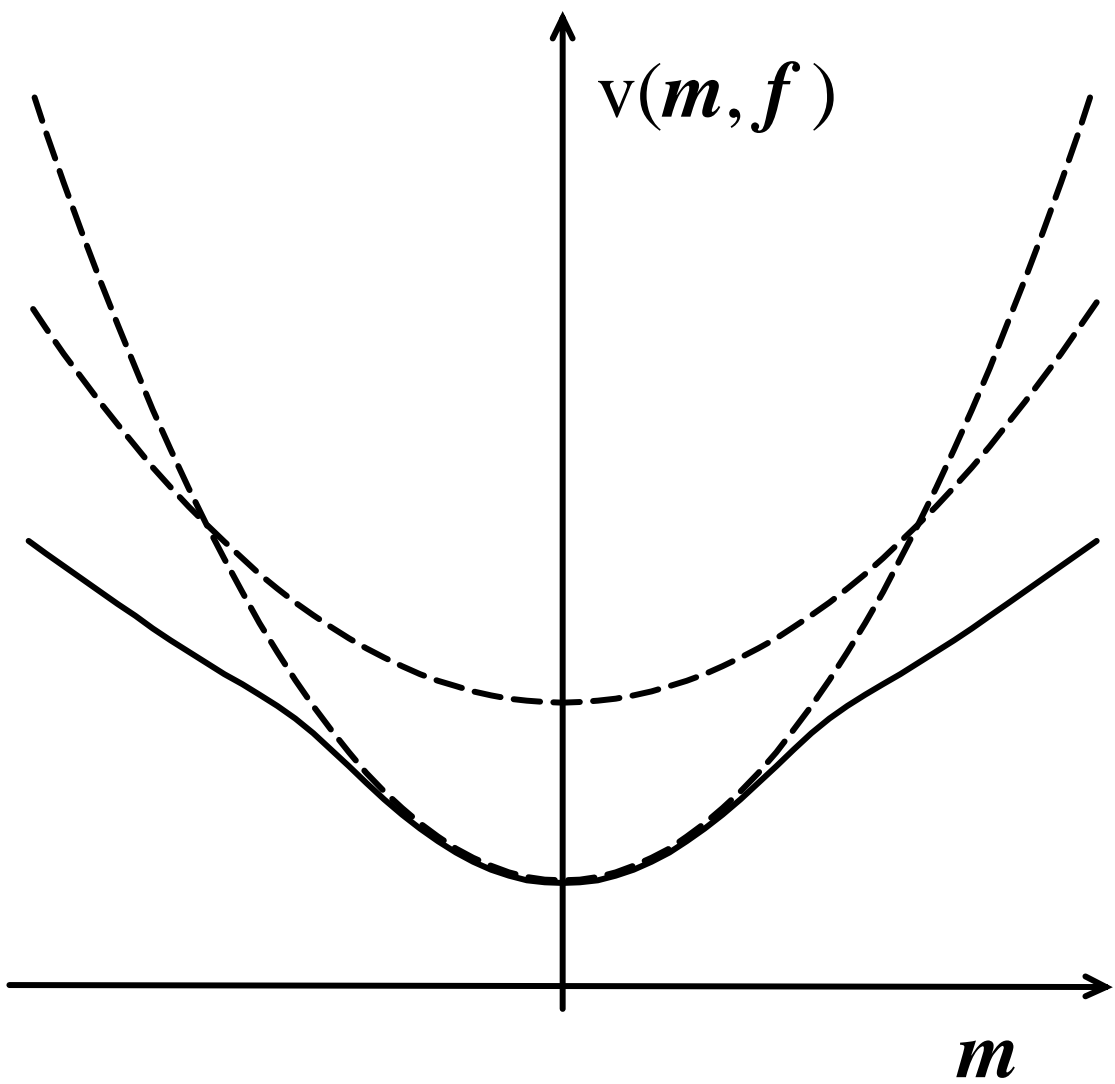
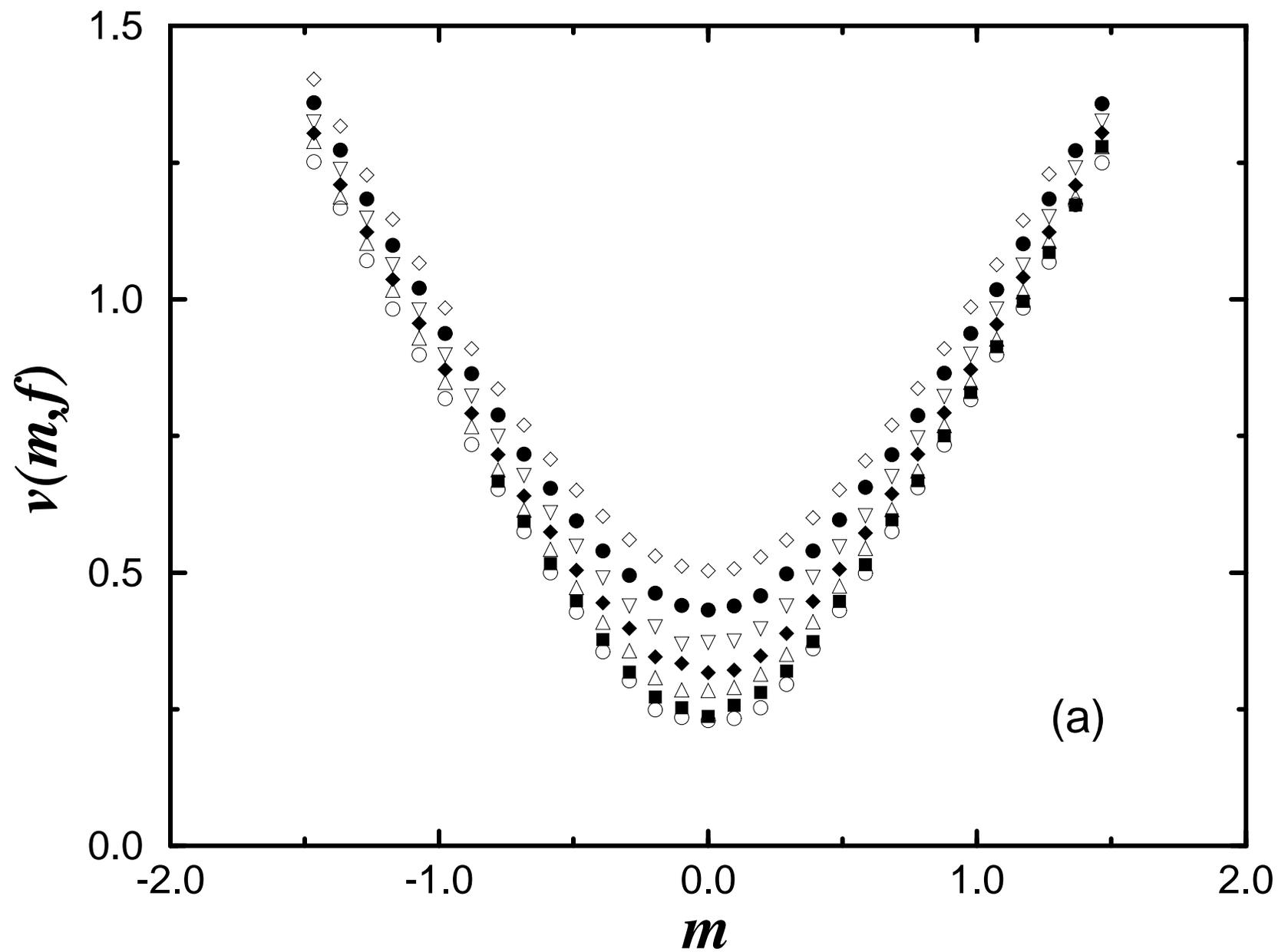
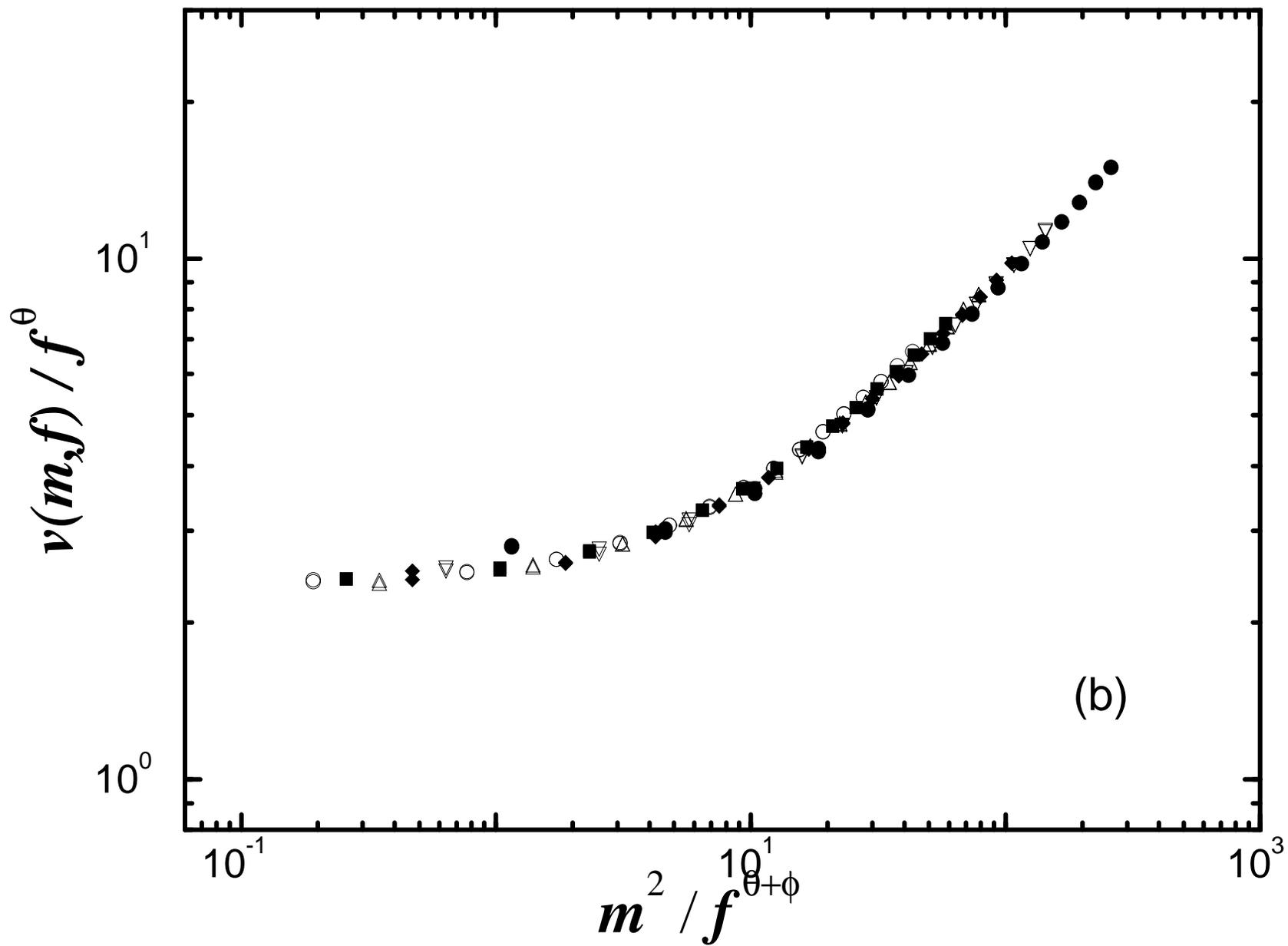
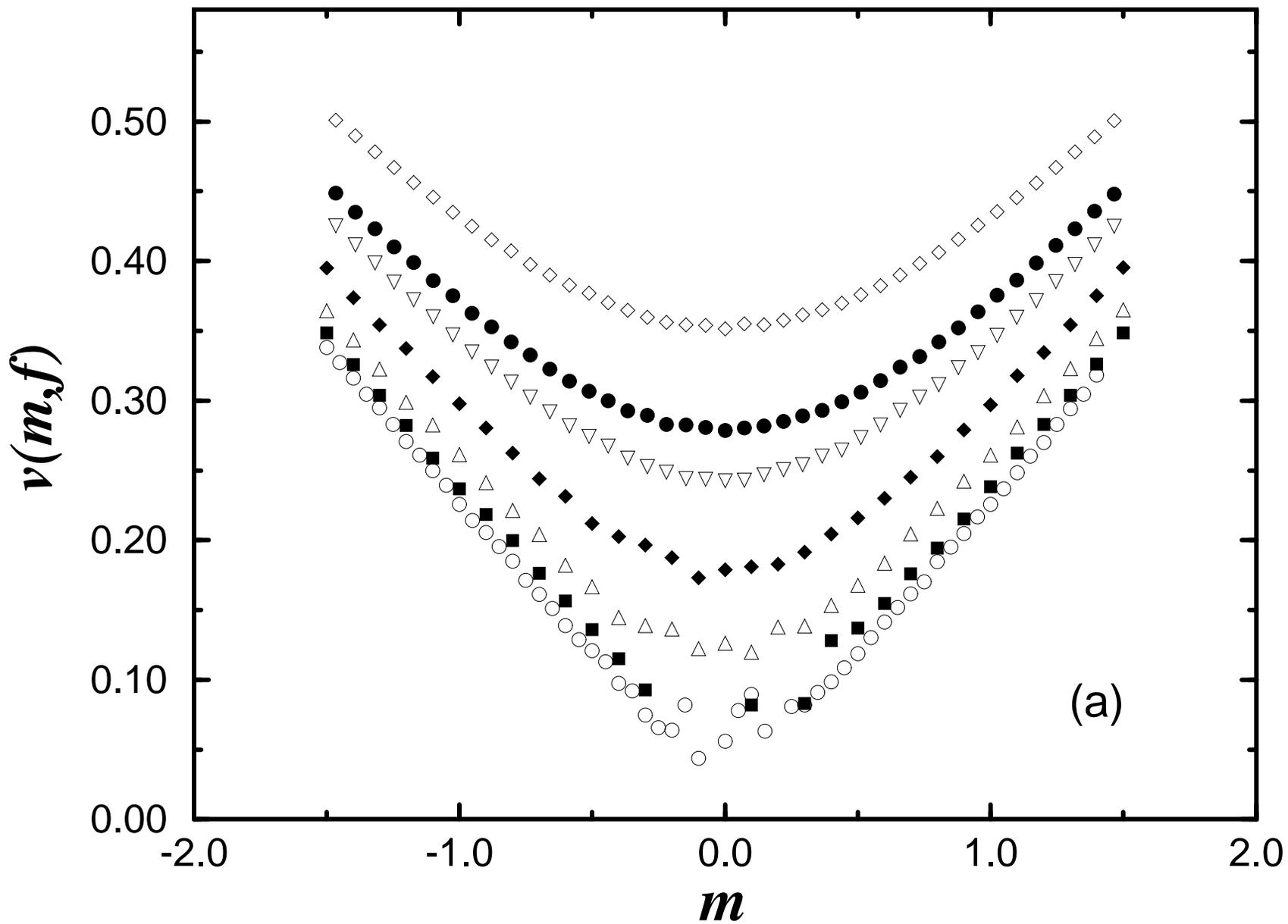
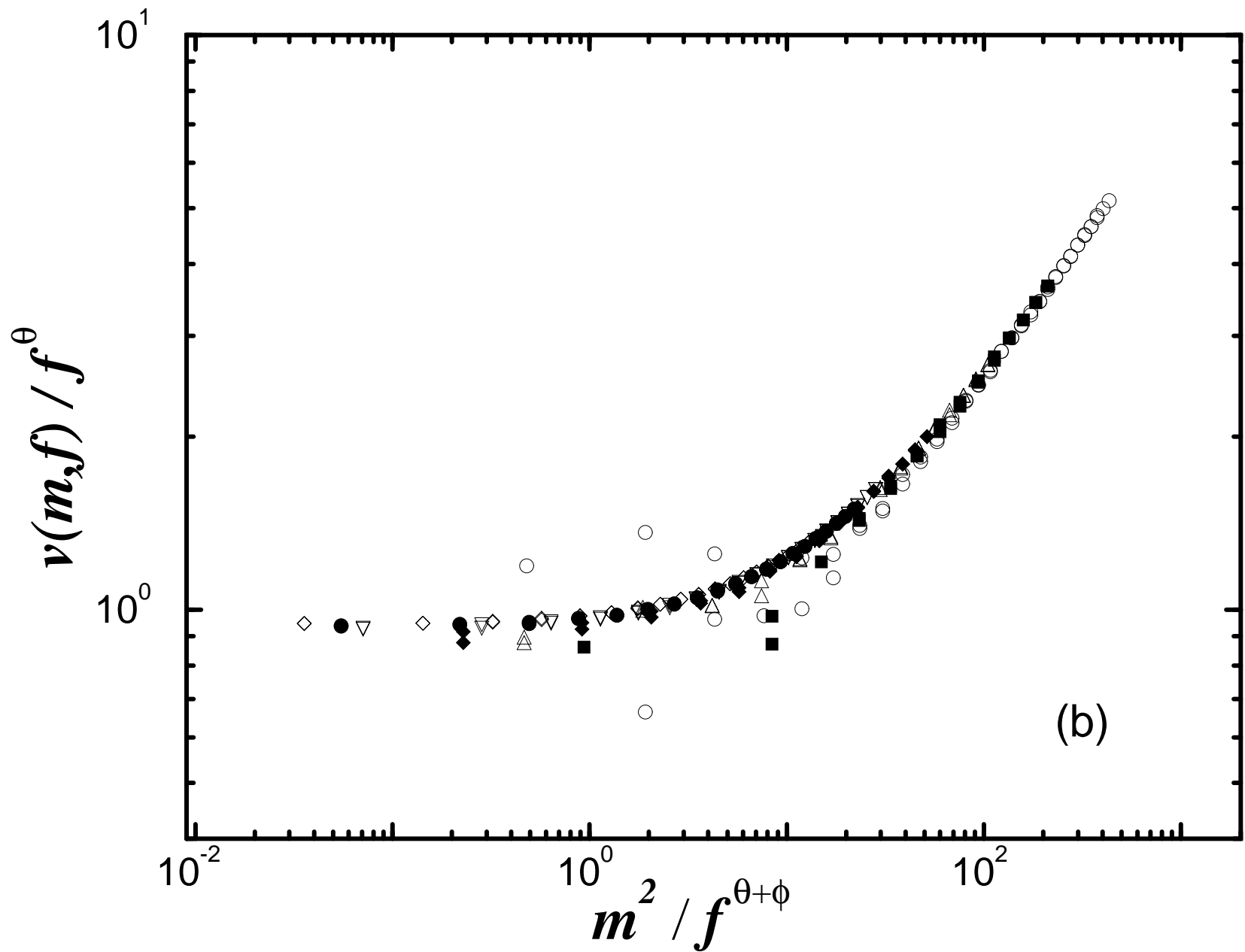
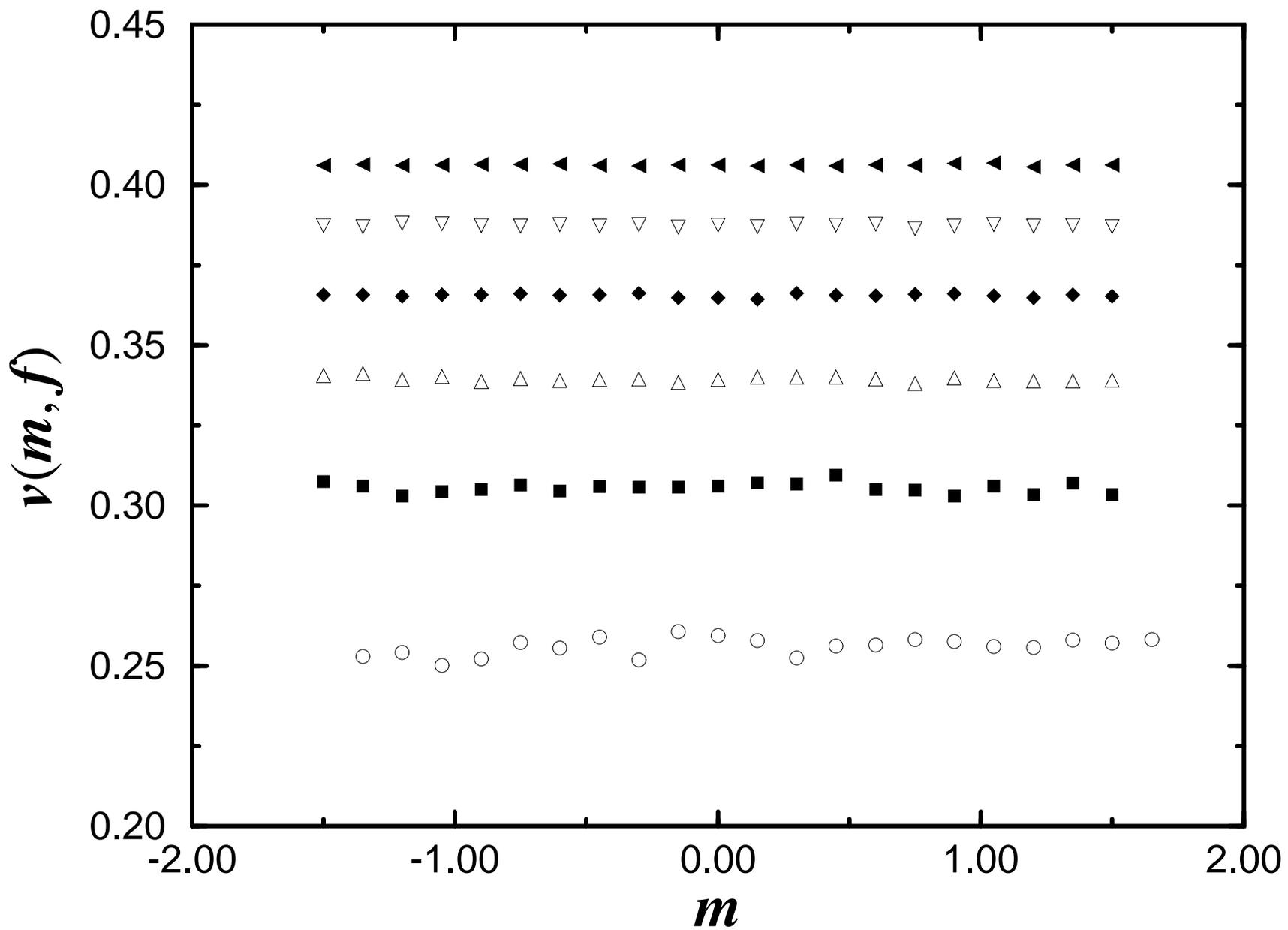
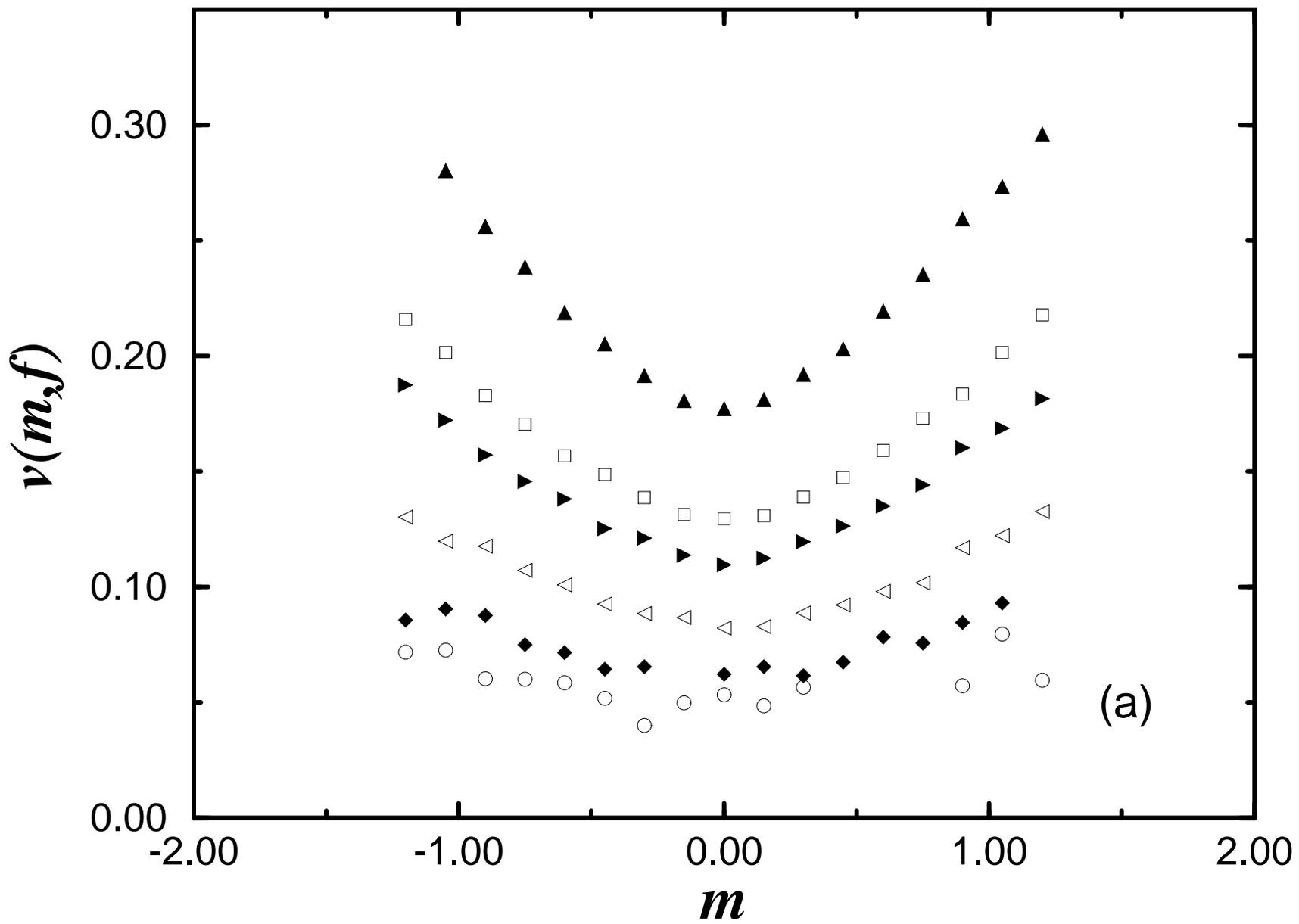
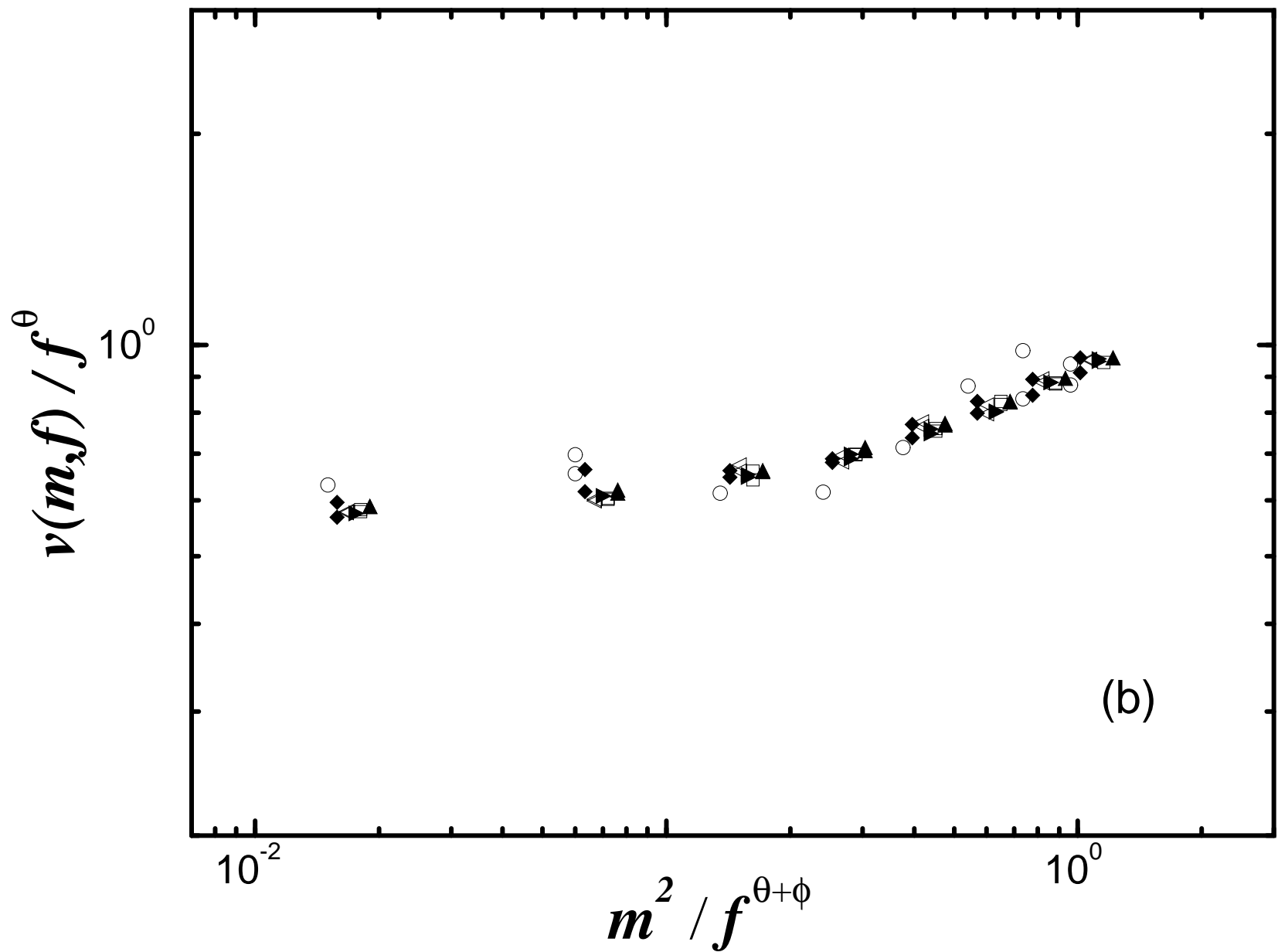
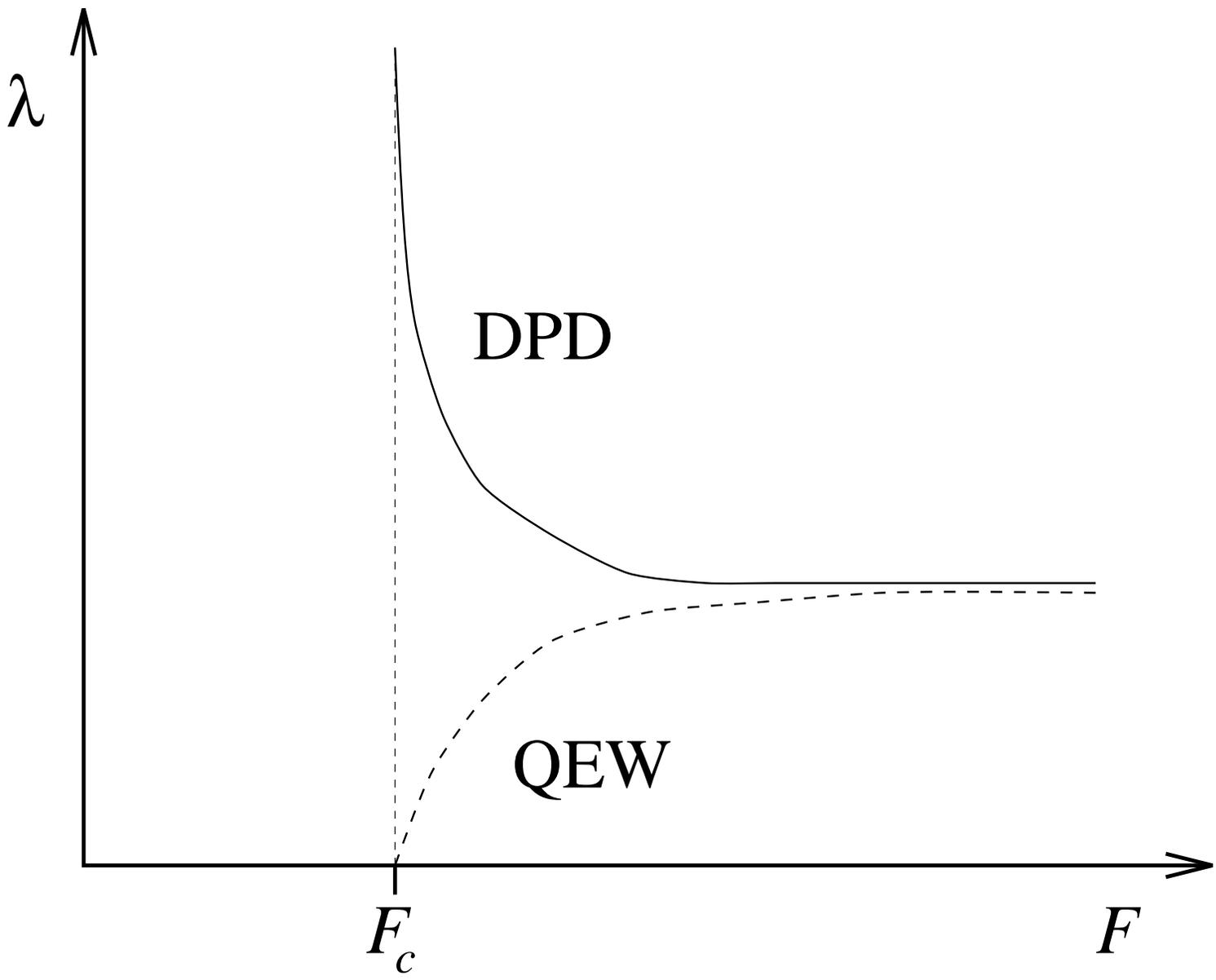
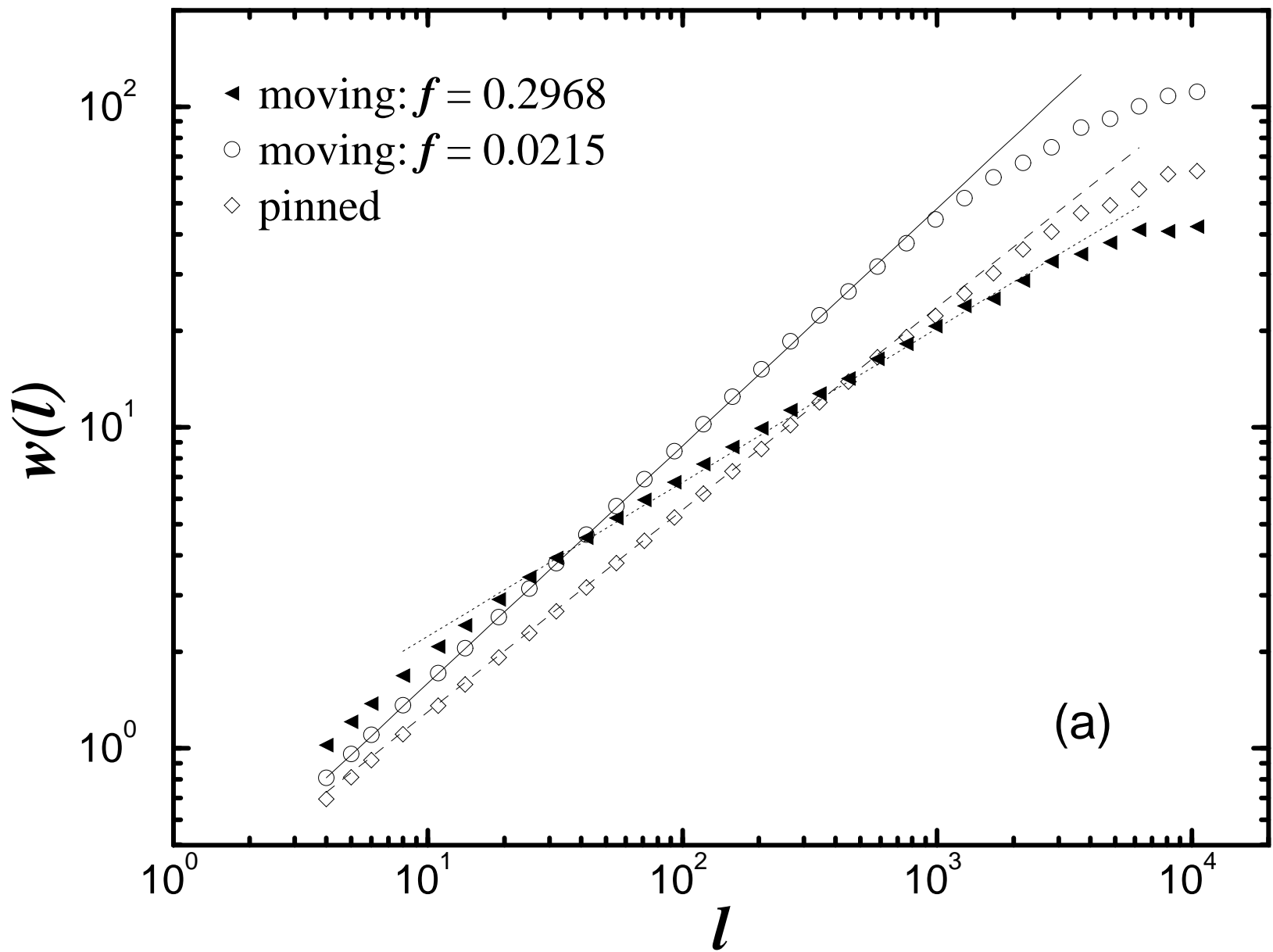
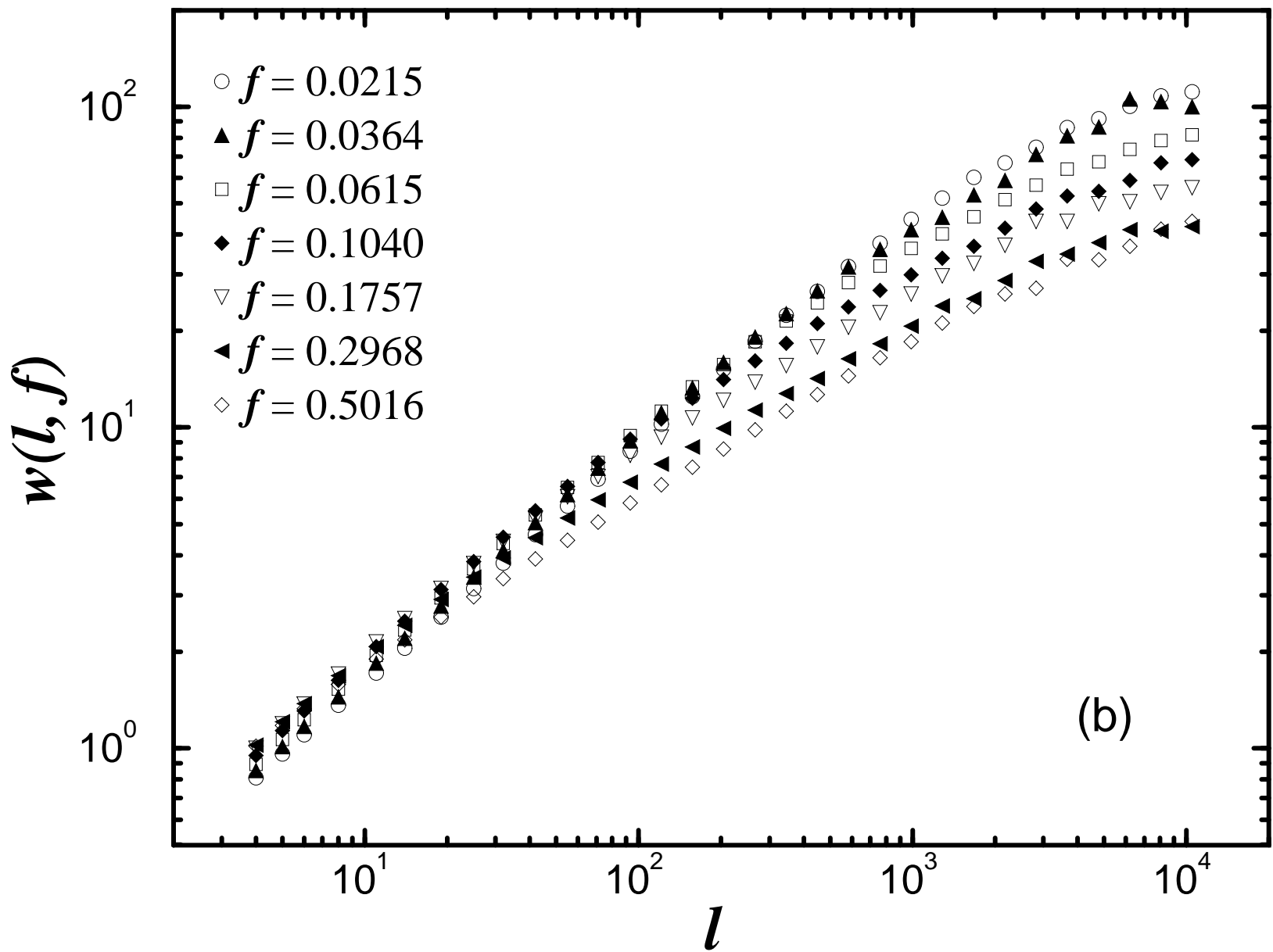
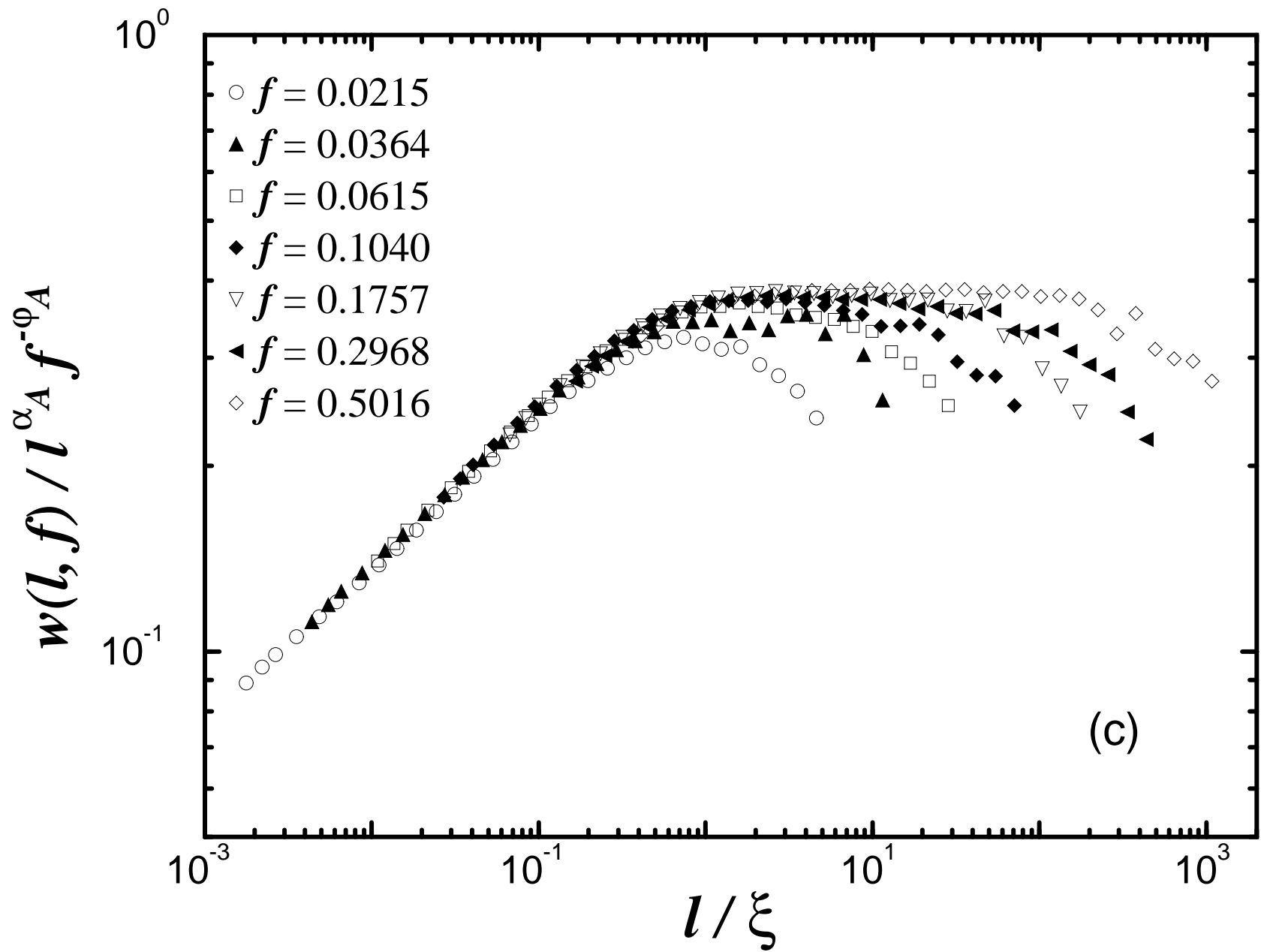
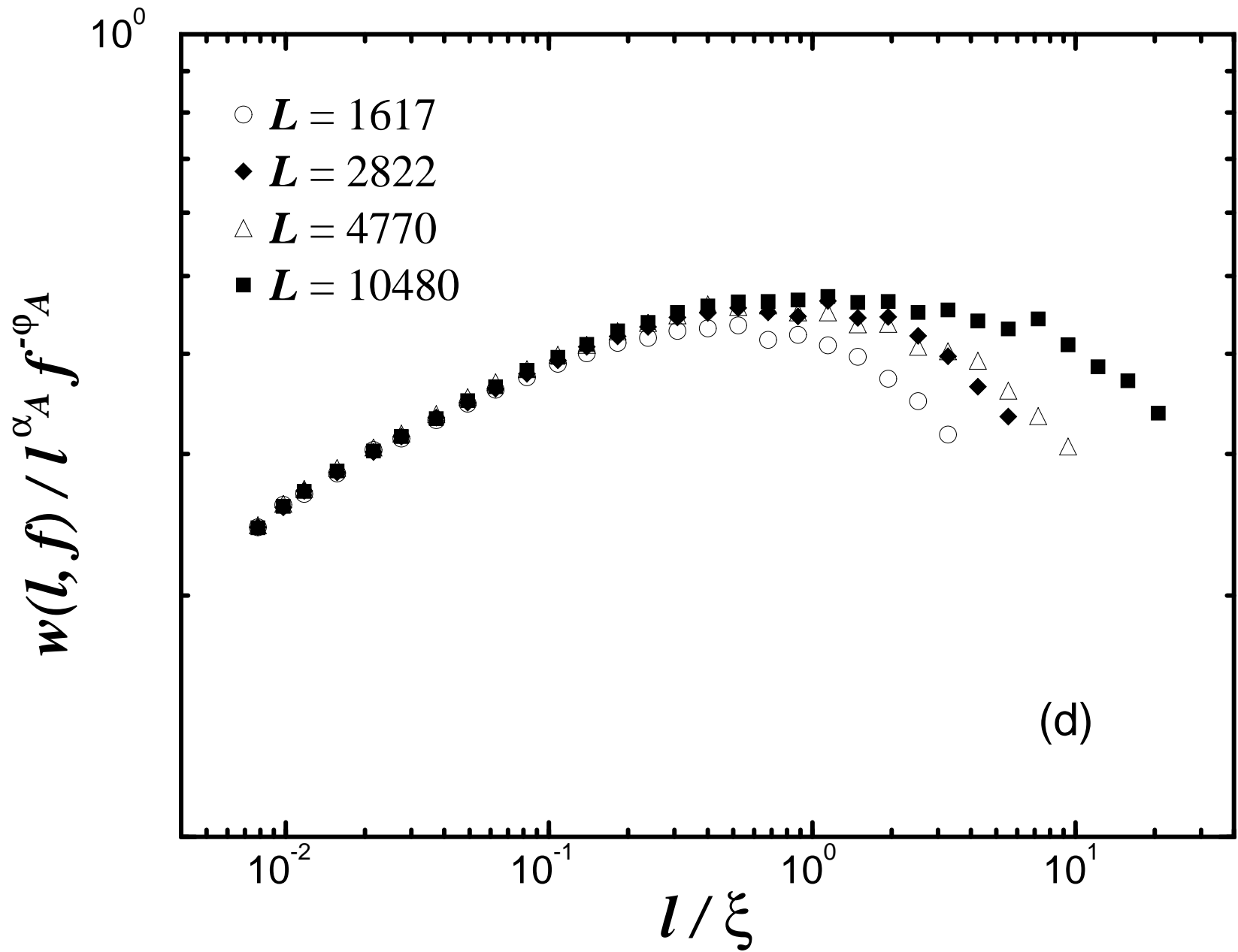
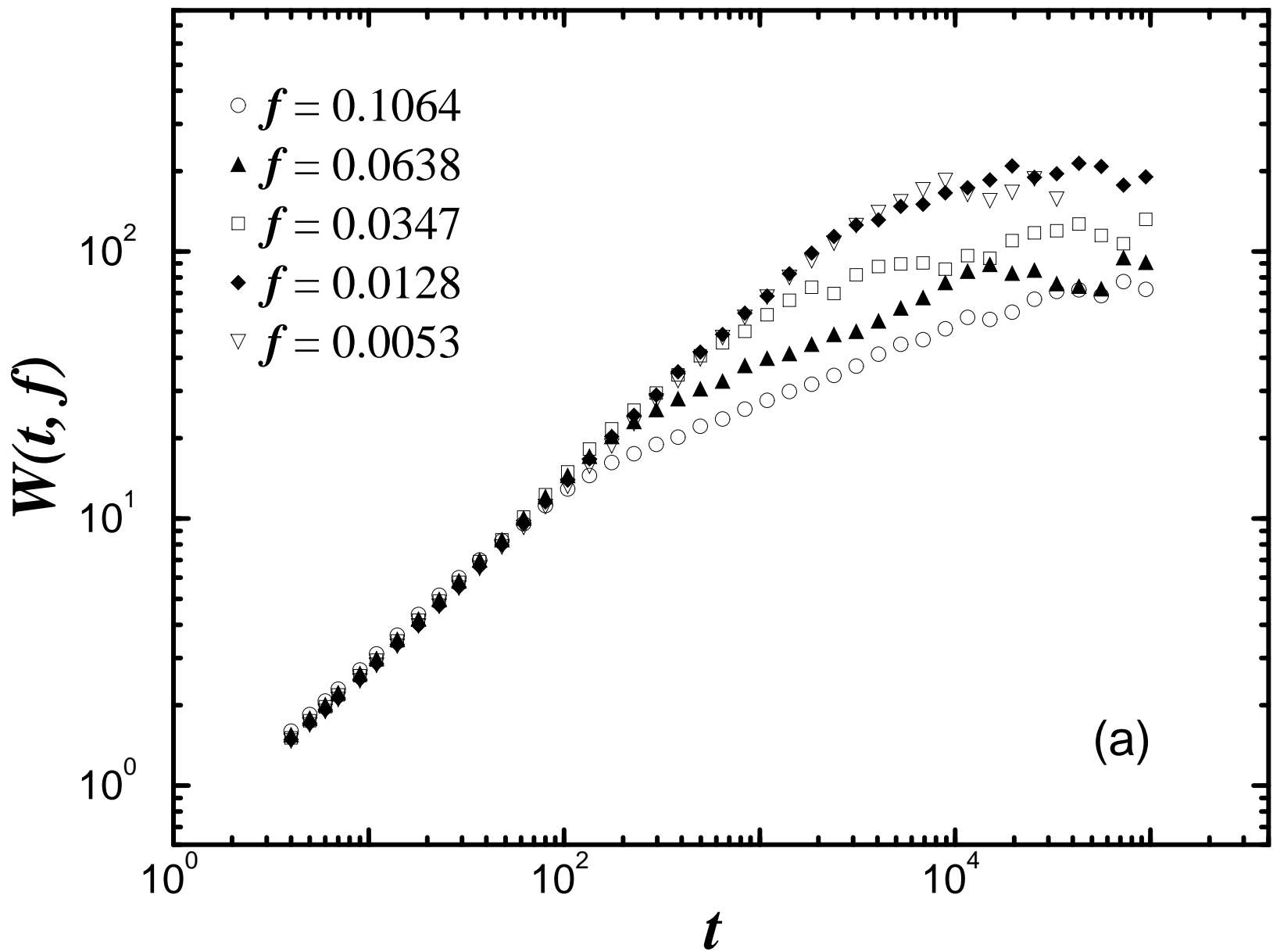
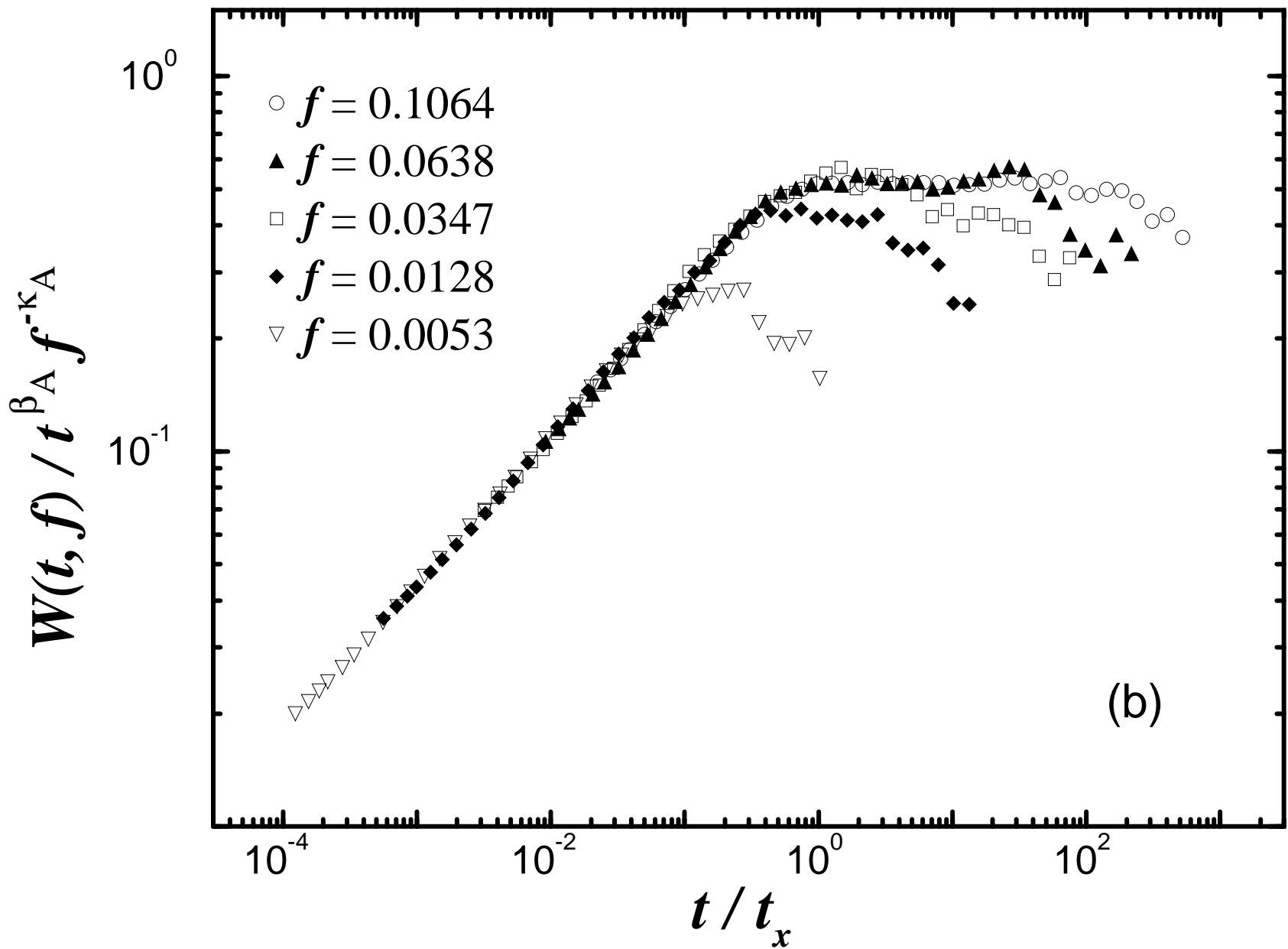
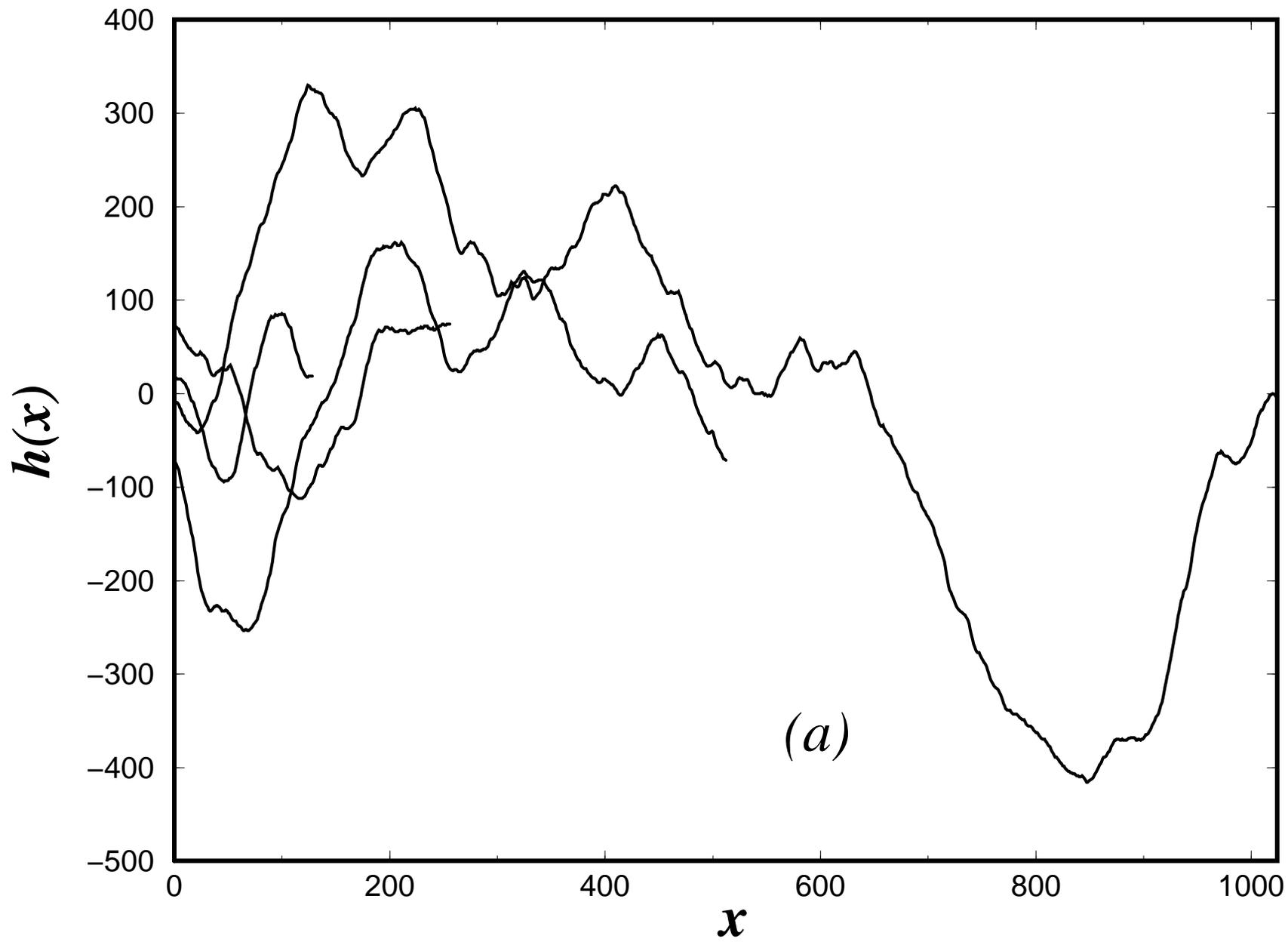
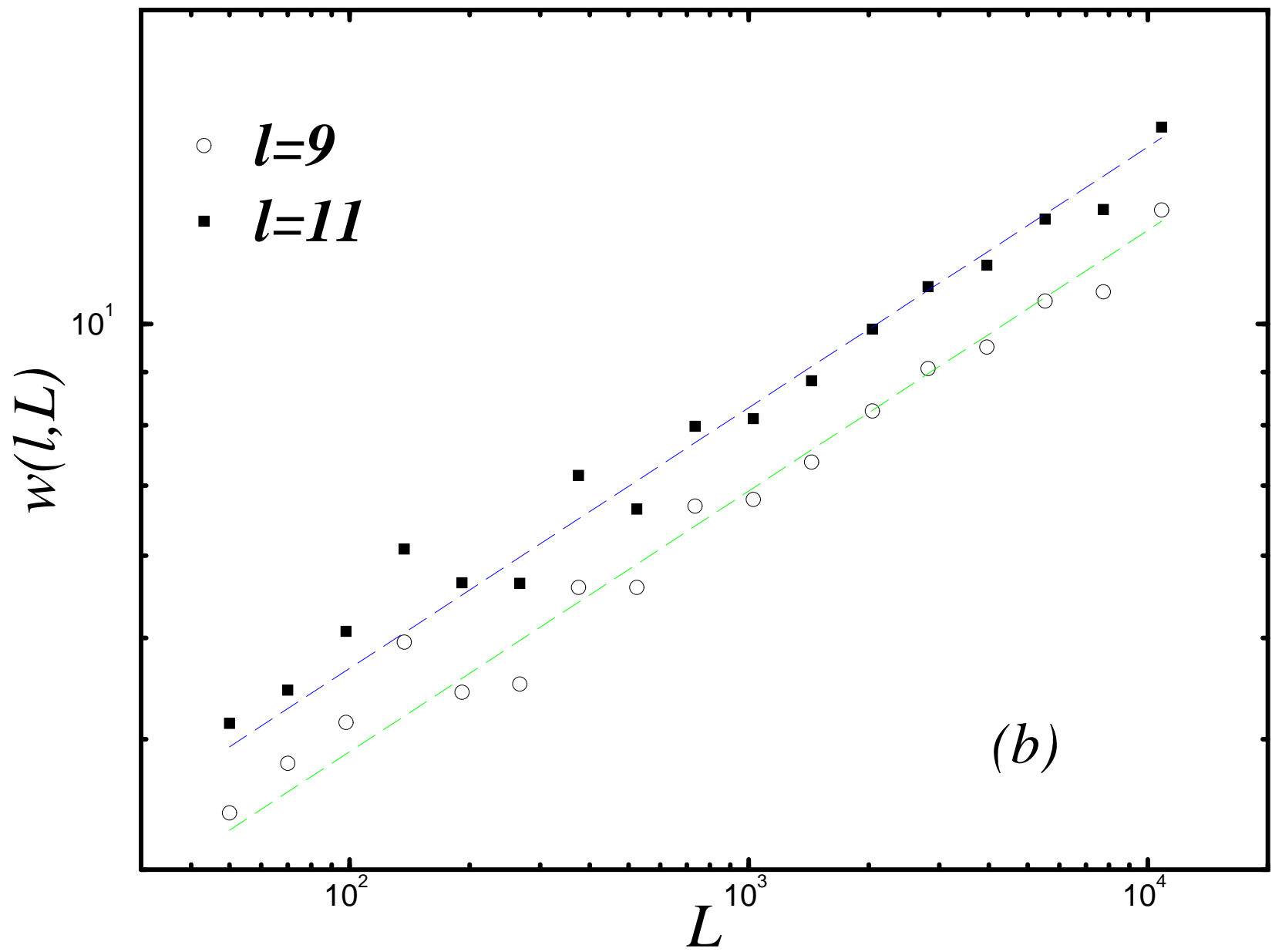
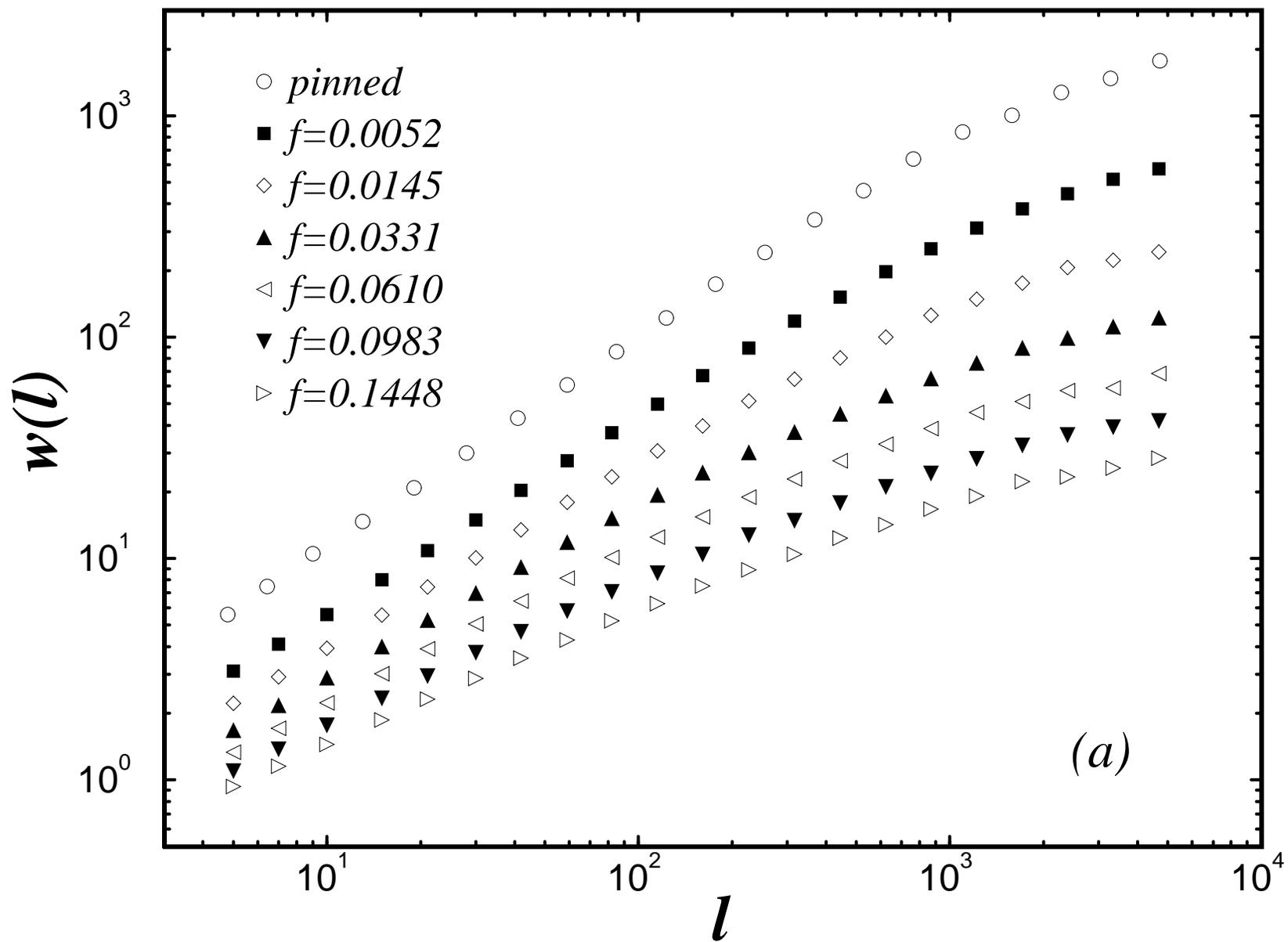
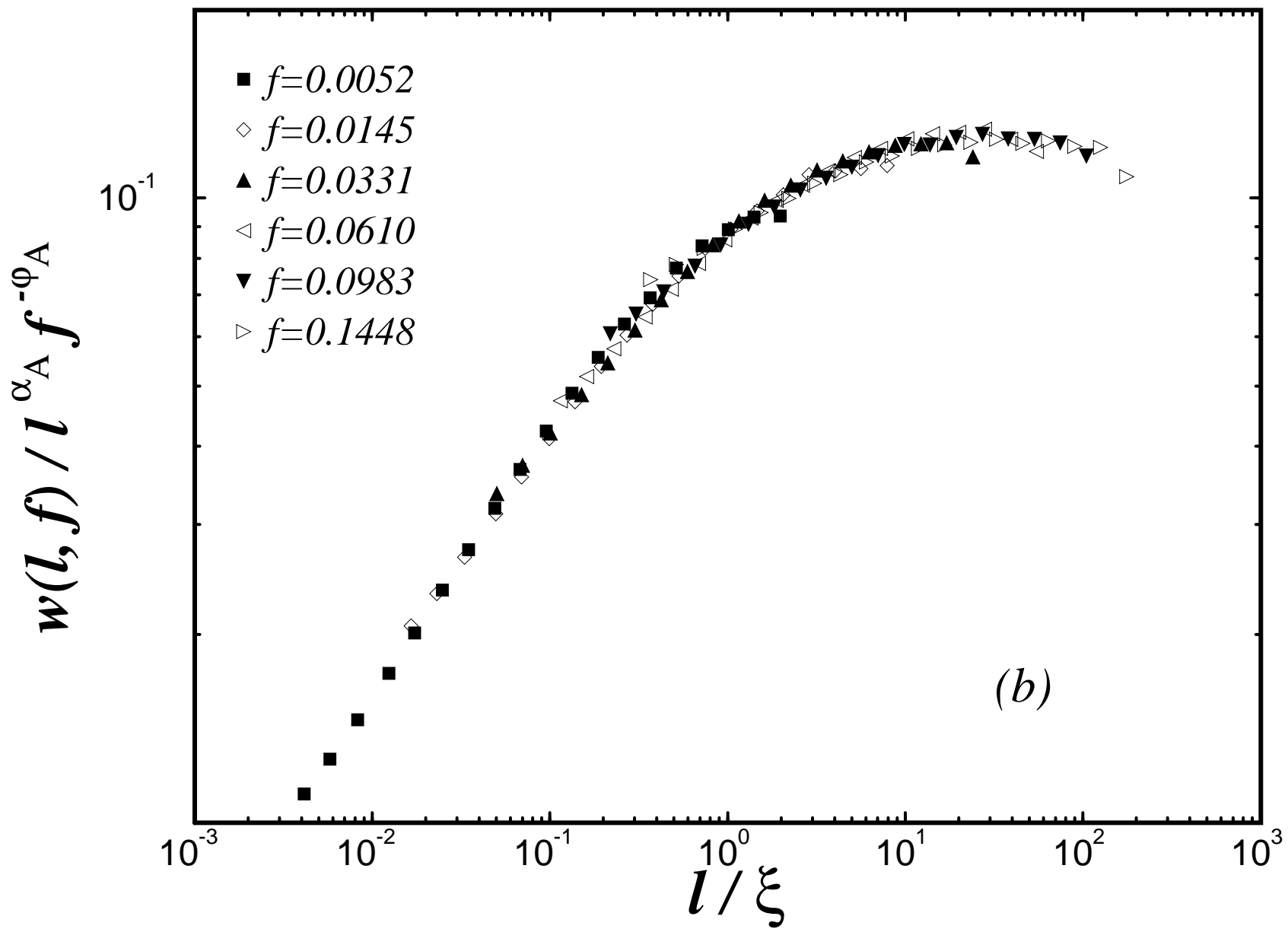
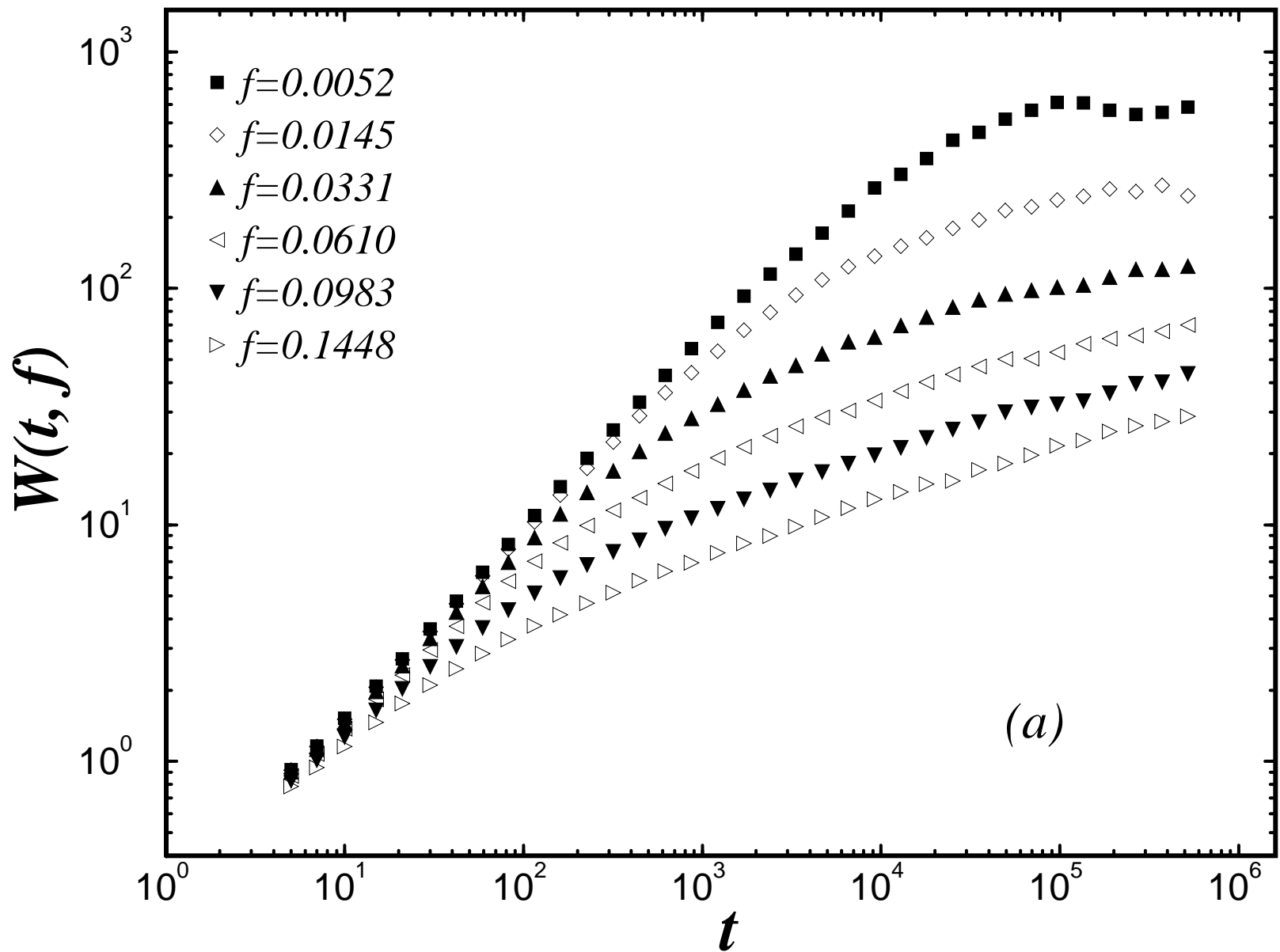
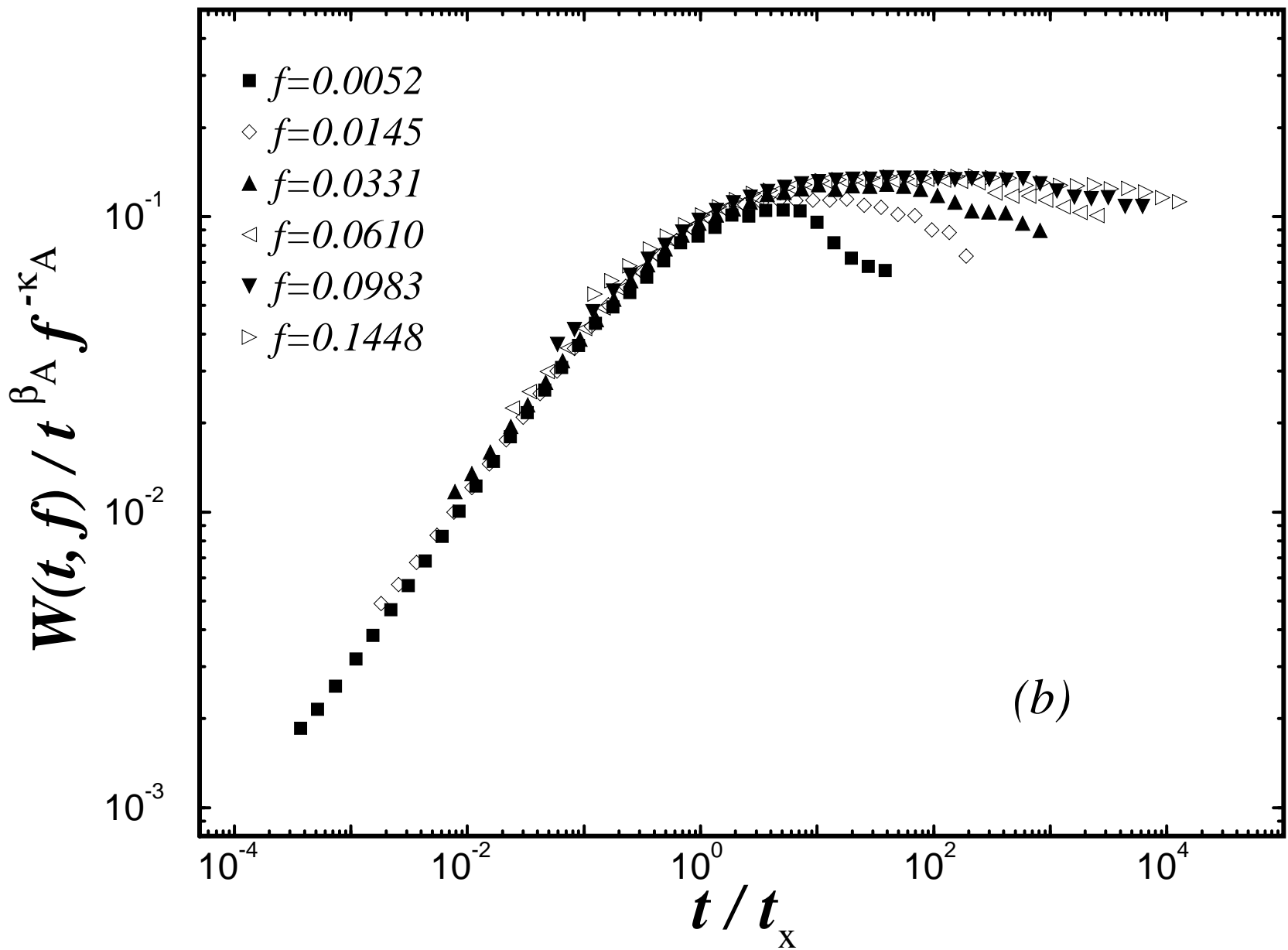
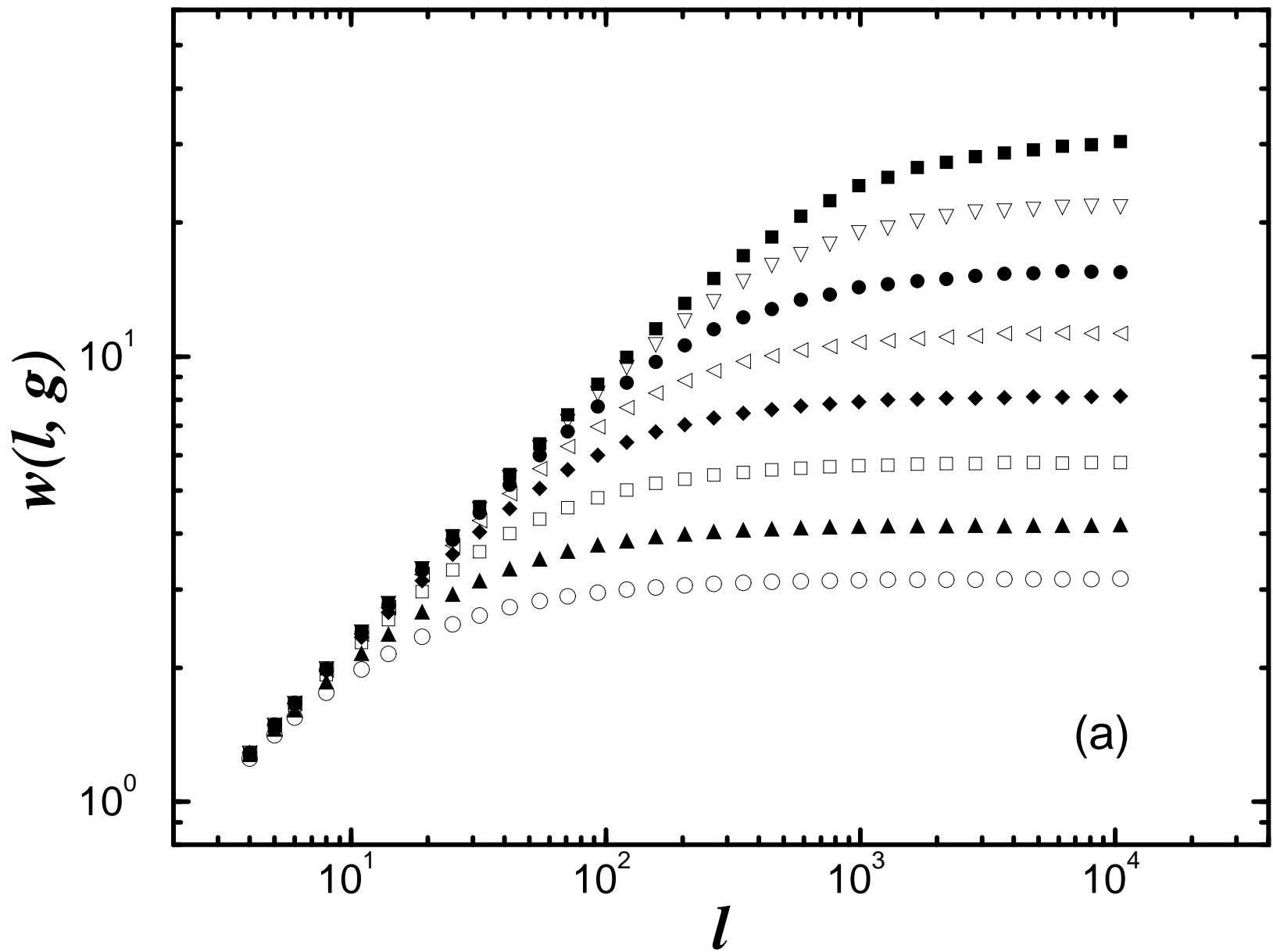
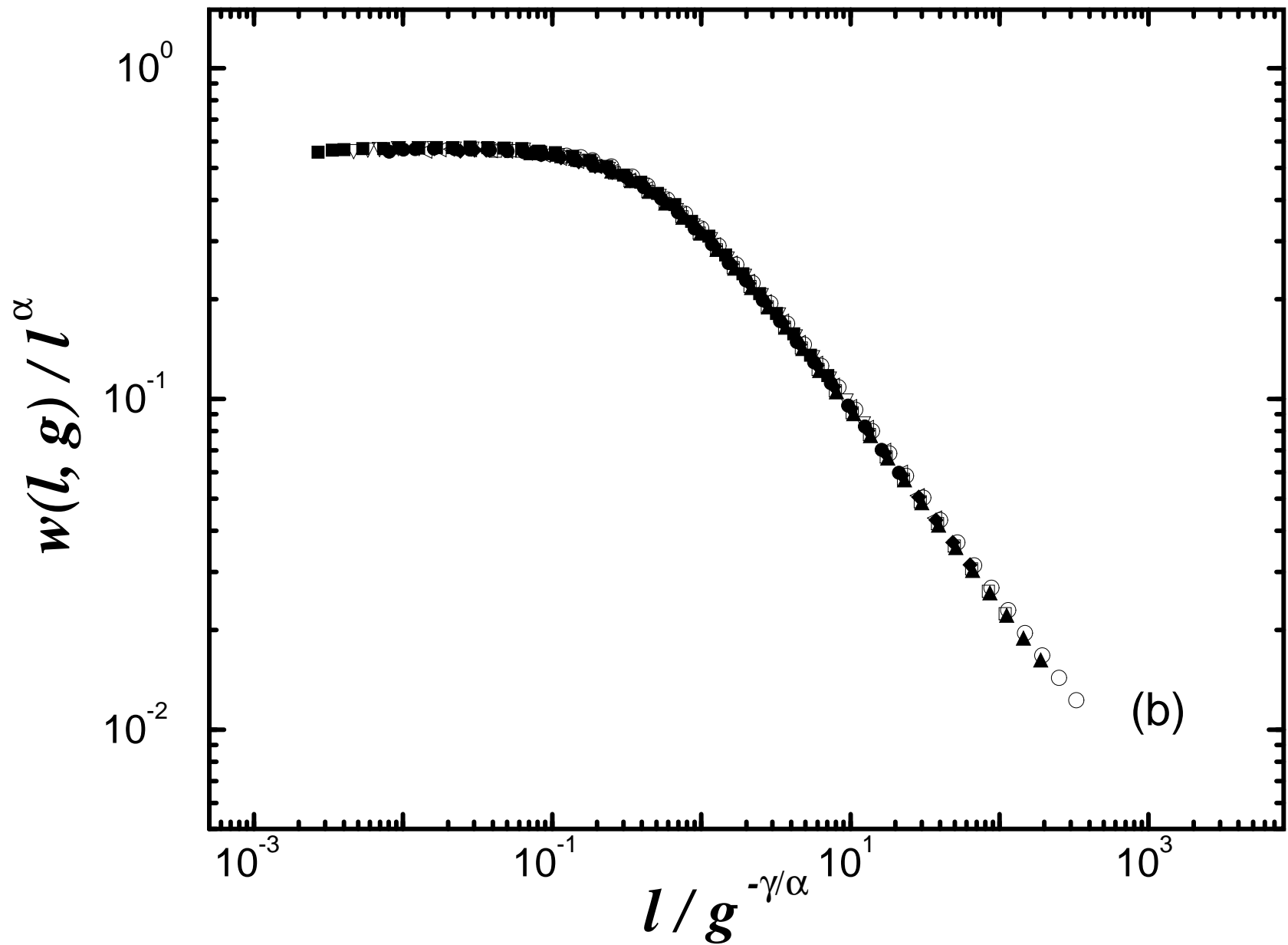
\end{document}